\documentclass[%
 reprint,
 amsmath,amssymb,
 aps,
 nofootinbib,
 longbibliography,
]{revtex4-1}

  \usepackage[pdftex]{graphicx}
  \DeclareGraphicsExtensions{.pdf,.png}

  \usepackage{mathptmx}
  \usepackage{amstext}
  \usepackage{amsfonts}
  \usepackage{IEEEtrantools}
  \newcommand{\beq}{\begin{IEEEeqnarray}{rCl}}
  \newcommand{\eeq}{\end{IEEEeqnarray}}

  \newcommand{\xcor}{s}

  \usepackage{soul}
  \usepackage{multirow}
  \usepackage{booktabs}
  \usepackage{longtable}
  \usepackage{ltxtable}
  \usepackage{tabularx}
  
  \usepackage[nomargin,inline,draft]{fixme}
  \fxusetheme{color}
  \usepackage{hyperref}
  \hypersetup{
    unicode=false,
    pdfcreator={pdflatex},
    pdfproducer={Latex with hyperref},
    pdfnewwindow=true,
    colorlinks=true,
    linkcolor=red,
    citecolor=green,
    filecolor=magenta,
    urlcolor=blue}

\begin{document}
  \preprint{APS/123-QED}

  \title{Quantifying power use in silicon photonic neural networks}

  \author{Alexander~N.~Tait}
  \email{atait@ieee.org}

  \affiliation{Physical Measurement Laboratory, National Institute of Standards and Technology,\\ Boulder, CO 80305, USA}





\begin{abstract}
  \noindent Due to challenging efficiency limits facing conventional and unconventional electronic architectures, information processors based on photonics have attracted renewed interest.
  Research communities have yet to settle on definitive techniques to describe the performance of this class of information processors. Photonic systems are different from electronic ones, so the existing concepts of computer performance measurement cannot necessarily apply.
  In this manuscript, we attempt to quantify the power use of photonic neural networks with state-of-the-art and future hardware. We derive scaling laws, physical limits, and new platform performance metrics.
  We find that overall performance is regime-like, which means that energy efficiency characteristics of a photonic processor can be completely described by no less than seven performance numbers.
  The introduction of these analytical strategies provides a much needed foundation for quantitative roadmapping and commercial value assignment for silicon photonic neural networks.
\end{abstract}



\maketitle

\section{Introduction}
  The computational requirements and, therefore, energy expenditures of machine learning are so staggering that its prevalence is becoming a climate issue~\cite{Strubell:19}. In the pursuit of energy efficiency, massively distributed hardware has been developed~\cite{Jouppi:17}. These pursuits have come to include non-digital signaling~\cite{Murmann:15a,Jain:20} and post-CMOS platforms~\cite{Sengupta:16}, including photonic platforms~\cite{Shen:17,Prucnal:17}.
  One of the value propositions of photonic neural networks and vector-matrix multipliers (VMM) is reduced energy use in performing linear operations.

\begin{figure*}[htb]
    \begin{center}
    \includegraphics[width=0.8\linewidth]{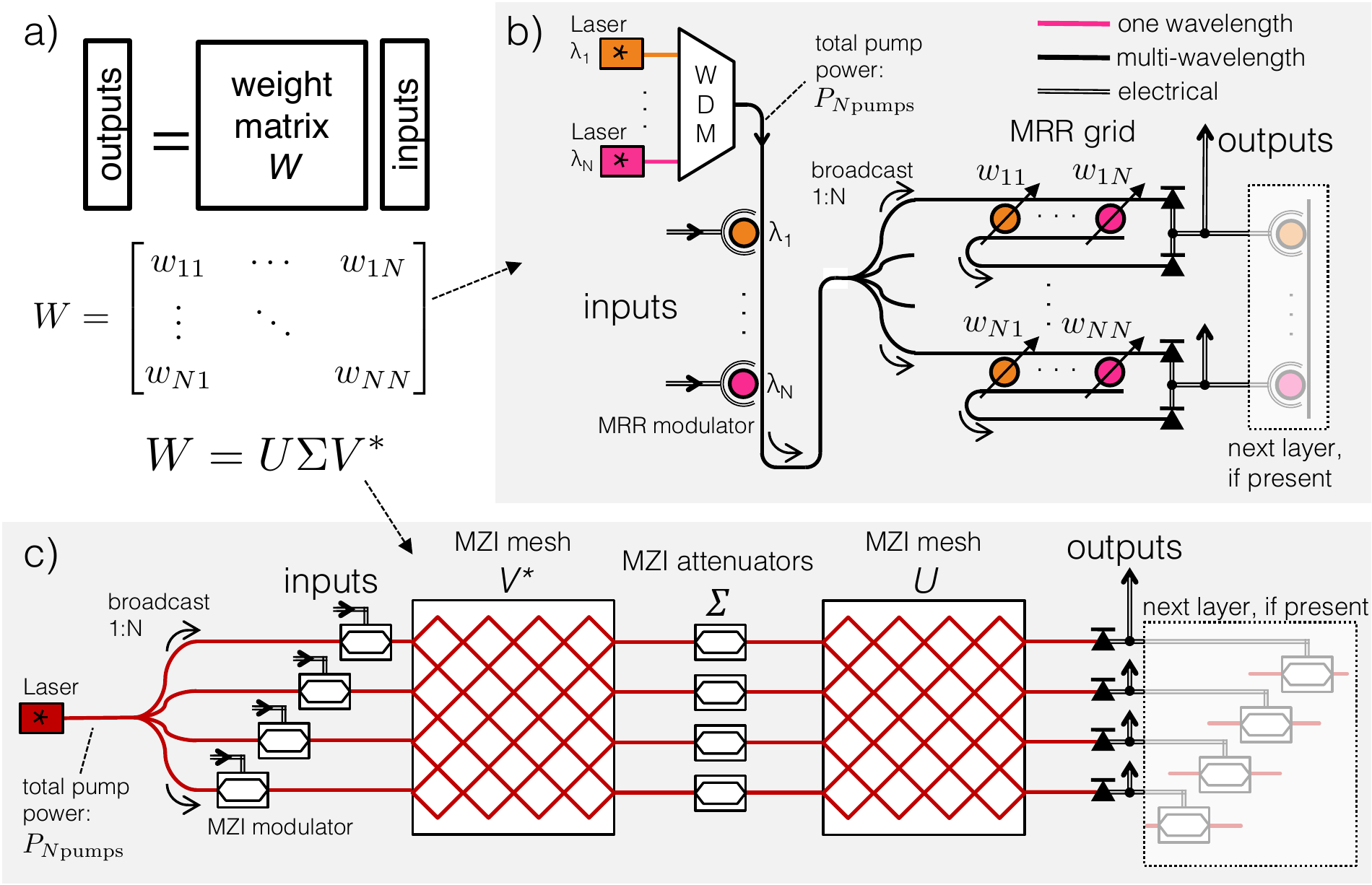}
    \caption[]{Silicon photonic neural network architectures. Optical pumps and electrical inputs/outputs are shown.
    a) A vector-matrix operation central to neural interconnects. The weight matrix, $W$, can be broken down into elements, $w_{ij}$, or into two unitary ($U$, $V^*$) and one diagonal ($\Sigma$) matrices.
    b) A multiwavelength broadcast-and-weight network~\cite{Tait:14}.
    Laser pumps are wavelength-division multiplexed (WDM). Each wavelength ($\lambda_1 \ldots \lambda_N$) is modulated by one electrical input, and all are broadcast. Microring weights in a grid layout each represent one element of the weight matrix. Balanced photodiodes detect the sum of each row. MRR color corresponds to the wavelength it acts upon.
    c) A silicon coherent nanophotonic network~\cite{Shen:17}.
    One laser pump provides power to every input beam. Each beam is modulated by one input. Tunable MZI meshes implement the unitary transforms. The singular, diagonal matrix corresponds to an element-wise multiplication, which is implemented by an array of attenuators. The optical output is converted back to the electrical domain. Depending on use case, the output can cascade to another layer (feedforward network, shown in dashed boxes), connect back to the original inputs (recurrent network), or merely be digitized and used elsewhere (vector-matrix multiplier). To function, both require some amount of optical power: $P_{N\text{pumps}}$. Low-bandwidth electronics for weight configuration are not shown. MRR: microring modulator; MZI: Mach-Zehnder interferometer.
    }
    \label{fig:conceptNetwork}
    \end{center}
  \end{figure*}

  It is essential to have a rigorous understanding of power scaling laws and limits in order to support that proposition. This understanding is foundational to system design and roadmapping, recognizing what the limiting factors are, and prioritizing technological developments that address those factors. In this manuscript, we study the energy consumption and computational efficiency of two silicon photonic neural network architectures~\cite{Tait:14,Shen:17} described in Fig.~\ref{fig:conceptNetwork}. We attempt to answer the key questions of how efficient they are in the worst case, how efficient they could be, and which technologies have the greatest impact on getting there.

  Energy efficiency of photonic neural networks has been studied in prior works in depth~\cite{Nahmias:20,Totovic:20}.
  Other works have proposed scaling laws that are only applicable to specific regimes~\cite{Shen:17,Williamson:20,Tait:17,Hamerly:19}. Some of these scaling laws have been extrapolated with gratuitous optimism to make predictions that idealized optical system can perform MACs for free in the limit of large matrices. We refute these predictions using energy conservation arguments. Large matrices are subject to different scaling laws that dictate that MAC efficiency approach finite values. This finding is true for both architectures analyzed: multiwavelength based on WDM weight banks and coherent based on Mach-Zehnder interferometers. These two architectures, despite distinct theories of operation (Fig.~\ref{fig:conceptNetwork}), are found to share several identical scaling laws.

  Despite what might be seen as pessimistic performance predictions (relative to past work), we find that photonic neural networks and VMMs can be highly competitive compared to state-of-the-art electronics. That being said, improving energy efficiency is not the only value proposition of photonic neural networks.
  Their bandwidth and latency can enable new real-time applications that are unaddressable by foreseeable electronic processors. The scope of this manuscript is only power use, not intended to minimize the pivotal role of quantitative studies of bandwidth, latency, and real-time applications.


  This manuscript takes a strategy of identifying invariant quantities that grant insight into the interplay of dominant power contributors.
  Total power use can be described as a sum of polynomials of the form $P \propto E N^x f^y z^B$. $N$ is number of channels, $f$ is bandwidth, and $B$ is resolution in bits -- these are termed functional parameters. Each polynomial term represents one power contributor.
  Power use ($P$) is described by scaling polynomials ($x$, $y$, $z$) and and energetic scaling coefficient ($E$) that is invariant for each power contributor.
  The power contributors dominate in different regimes of ($N$, $f$). By deconstructing these regimes, this approach provides more detailed insight than approaches that simply calculates the overall power or approaches that do not account for all of the contributors.



  The novel aspects of this manuscript can be organized by section. Section~\ref{sec:weight-control} includes the first quantification of expected power to counteract fabrication variation in a square matrix of microring resonator (MRR) weights. We propose a path to reduce this power by 4 orders-of-magnitude by pairing two technologies.
  Sec.~\ref{sec:resolution-limited} derives resolution-determined scaling laws and closed-form expressions of their energetic scaling coefficients. Prior works have studied physical sources of noise in a single neuron~\cite{Lima:20} and network scaling behavior in an abstract sense~\cite{Semenova:19}.
  To the author's knowledge, this is the first physics-based derivation of resolution concepts in any type of multi-channel photonic information hardware~\cite{Bogaerts:20}. Previously unrealized insights about noise in photonic information processing result, including a physical limit on MAC efficiency and a hard limit on bandwidth.

  Sec.~\ref{sec:gain-limited} analyzes gain-determined power as set by cascadability and/or digitization requirements. Arguments based only on energy conservation and functional generality results in a scaling law that refutes free-lunch notions presented in numerous prior works. Sec.~\ref{sec:optoelectronic-switching} examines O/E/O transduction in analog neurons and proposes a role for photoelectric amplifiers.
  Scaling laws and energetic coefficients relating to noise, gain, and detection are found to be nearly identical between multiwavelength MRR-based (Fig.~\ref{fig:conceptNetwork}b) and coherent MZI-based (Fig.~\ref{fig:conceptNetwork}c) architectures for both photonic neural networks and VMMs. They are described in terms of the same set physical variables. These similarities are surprising because of the fundamental differences in how the architectures employ different properties of light.
  Section~\ref{sec:summary} compares all of these contributors in terms of dominant regimes, in the process providing a roadmap for device technologies and a walkthrough of thought processes for future system design.

\section{Weight Control} \label{sec:weight-control}
  Photonic weighted addition schemes depend on the ability to configure the transmission state of passive elements. All proposals for programming weights so far employ thermooptic tuning to achieve the necessary phase shifts. The required power breaks down into a static and a configurable component and is proportional to the number of weights.
  \beq
    P_{wei} &=& N^2 \cdot \left(P_{lock} + P_{conf}\right)
  \eeq
  The static component is needed to lock MRR weights onto their resonances, counteracting fabrication variations. The configurable component is the power needed to tune the MRR on and off resonance in order to program the desired weight value. When MRR modulator neurons are used, these same principles apply to the neurons. We will leave out this contributor below because it scales linearly with number of neurons, so it will not contribute as much as weights.

  \subsection{Weight locking power}
    Weight locking power is the electrical power needed to bias a weight. MZI architectures do not need biasing because MZIs are wavelength independed. MRRs, on the other hand, must be held close to the on-resonance condition with a WDM carrier. Fabrication nonidealities result in a wide variability in fabricated resonant wavelength. The only tuning effects strong enough to counteract this variability are thermal. Thermal locking dissipates a large and static amount of heat on chip.
    In Ref.~\cite{Narayana:17jrnl}, it was found that, in some operating regimes of a photonic network-on-chip, static MRR heating accounts for up to 80\% of total power. Static locking can dominate in simple communication links, such as in Ref.~\cite{Zheng:14} (80\% of total), but not always~\cite{Timurdogan:14} (10$^{-4}$--23\% depending on temperature).

    Over the chip area, a given resonance can vary more than an FSR from fabrication target, given current fabrication abilities. At most one FSR of tuning range is needed to put one resonance onto a given wavelength target. The standard deviation of resonance offset is correlated with distance, such that nearby MRRs are likely to vary relatively little from one another as compared to their absolute variation. For resonators spaced by $r$[mm], total standard deviation is
    \beq
      \sigma_{[FSR]}(r) = \sigma_{[FSR]}^{(0)} + \sigma_{[FSR]}^{(1)} r
    \eeq
    where the subscript $[FSR]$ means in free spectral range units. Reference~\cite{Chrostowski:14} measured these parameters for the IME A*STAR process to be $\sigma_{[FSR]}^{(0)} = 0.050$ and $\sigma_{[FSR]}^{(1)} = 0.060$mm$^{-1}$ with the FSR at 7nm in that work. Variances can also be stated in wavelength units (denoted with subscript $[\lambda]$) by multiplying by the FSR in wavelength units.

    We introduce a term $\Omega$ to indicate the expected value of resonant shift per MRR needed to bring a square array of MRRs onto resonance.
    \beq
      \Omega(N) &=& \min\left[\sigma_{[FSR]}(Nd), \frac{1}{2}\right] \\
      P_{lock} &=& K \Omega(N) \label{eq:locking-power}
    \eeq
    where $d$ is the MRR pitch, and $Nd$ is the side length of the square MRR array. $\Omega$ is in FSR units. $K$ is the tuning efficiency in mW per FSR units. The MRR pitch, $d$, is taken here to be 20~$\mu$m.

  \subsection{Weight configuration power}
    Weight configuration power is used to program the weight value. Applying heat tunes the MRR from on-resonance (weight --1) to slightly off-resonance (weight +1).
    Supposing that tuning over a full-width half maximum (FWHM) is required, this power is

    \beq
      P_{conf} = \frac{K}{2\mathcal{F}} \label{eq:configuration-power}
    \eeq
    where $\mathcal{F}$ is finesse. We state $K$ in FSR units, so finesse converts it to FWHM units. The factor of 2 results from averaging over the range of possible states from on-resonance to off-resonance by one FWHM. We can approximate finesse as roughly 100 for typical silicon MRRs, although optimized traveling wave resonators have achieved finesse up to 1140~\cite{Soltani:10}. The vertical junction depletion modulators discussed below had a finesse of 277. Typical values for $K$ are given in~\ref{tab:weight-locking}.

    The MRR resonance has a sharp wavelength dependence, meaning that -- once locked -- there is a small incremental power needed to configure the weight. If we are unable to control where the resonance falls as fabricated, then locking power will be greater than configuration power by a factor of the finesse.



    \subsubsection*{MZI weight configuration}
      The opposite power balance is found in MZI mesh architectures. MZIs are less sensitive to fabrication variation, and they are correspondingly less sensitive to desirable tuning effects. MZI tuning power is quantified by $\pi$-power,~$P_{\pi}$:~the power needed for a thermal phase shifter to impart an optical phase shift of $\pi$. In the above architecture from~\cite{Shen:17}, there are four phase shifters per matrix element whose average expected power is halfway between minimum phase ($0$) and maximum phase ($\pi$) states.
      \beq
        P_{conf, MZI} = 2 P_{\pi, MZI} \label{eq:mzi-power}
      \eeq
      MZI thermal $\pi$-powers are on the order of of 10mW~\cite{Annoni:17}. Prior work on MZI meshes has calculated system power to be $\sim 1mW \cdot N$~\cite{Shen:17} or $\sim 100mW \cdot N$~\cite{Williamson:20}, but then the neglected weight configuration power, which would severely dominate at $10mW \cdot N^2$.

      MZI meshes do not need static weight locking power; however, MZI configuration power and MRR locking power are both in the mW range. They stem from different needs. The first is due to an essential need to program the weights, and the second is due to fabrication non-ideality.
      This means that MZI meshes have a fundamental need for strong tuning effects, while MRR weights can reduce tuning power by addressing fabricated resonance variability.

  \subsection{Weight reconfiguration energy}
    Weight reconfiguration energy is additional energy needed to change weights on a fast timescale, which is distinct from weight tuning. Typically, fast tuning reqires non-thermal tuning, such as depletion modulators that do not draw continuous power.
    Reconfiguration energy is

    \beq
      P_{reconfig} = N^2 f_{reconfig} \cdot E_{reconfig}
    \eeq
    where $E_{reconfig}$ is the same as the energy-per-bit value when considering each tuning element as a modulator. The reconfiguration rate, $f_{reconfig}$, is an expected, averaged rate that is less than the maximum reconfiguration bandwidth. $f_{reconfig}$ is highly application dependent.

    In terms of technology, larger tuning elements usually have higher capacitance and consequently higher reconfiguration energy. At the same time, those devices with higher capacitances -- or, in the case of MEMS, mechanical timeconstants -- have lower maximum switching frequencies. The higher switching energies and longer switching times of large devices have opposing effects on this power contribution.

    In many neural networks, the reconfiguration of weights happens at much slower timescales than signal timescales. In those cases, the power needed to change weights can generally be neglected. In other applications, such as general VMMs, weights must change at timescales similar to the signals. Since it is application-dependent, we largely leave reconfiguration energy from the rest of the analysis.

  \subsection{Foreseeable technology} \label{sec:foreseeable-tuning}
    Resonator locking metrics are listed in Table~\ref{tab:weight-locking}. We identify critical technologies that impact the locking power and configuration power: trench isolation for heaters~\cite{Dong:10opex, Cunningham:10}, photonic microelectronic mechanical systems (MEMS)~\cite{ErrandoHerranz:20}, resonator variability reduction~\cite{Alipour:15}, and/or interleaved junction modulators~\cite{Timurdogan:13,Timurdogan:14}.
    Thermal isolation trenches simply improve the thermal tuning efficiency by one order of magnitude. Several more orders of magnitude reduction could be realized with low-power, non-thermal tuning effects.

\begin{figure}[tb]
      \begin{center}
      \includegraphics[width=0.5\linewidth]{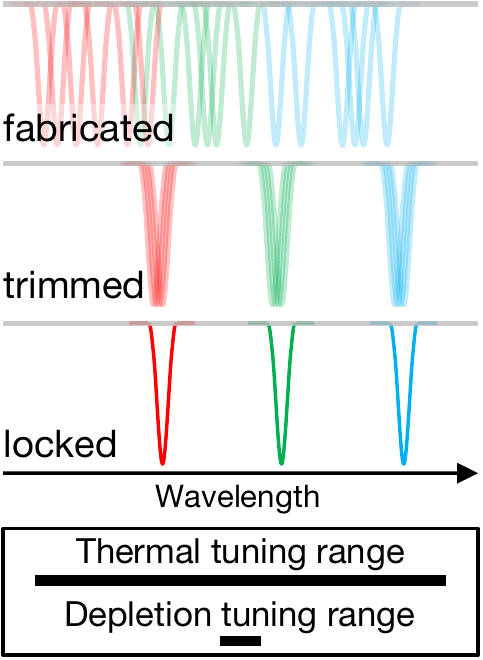}
      \caption[]{Concept of using trimming to reduce resonator variation below the depletion-mode tuning range.}
      \label{fig:trimming}
      \end{center}
    \end{figure}

    Tuning power can be reduced by approximately five orders-of-magnitude using one of two approaches. One -- applicable only to MZIs -- is using MEMS tuning. The MEMS approach would lead to weights that can be changed on the 100~kHz--10~MHz scale and would require a wet undercut etch sometime before metallization steps~\cite{ErrandoHerranz:20}. The other approach -- applicable only to MRRs -- is to combine low-power tuning with a variability reduction technique. This approach, introduced in Fig.~\ref{fig:trimming}, could be favorable because weak tuning devices are highly developed and already present on mainstream silicon photonic platforms.

    Both MZIs and MRRs benefit from strong, low-power tuning effects. Low-power tuning technologies can be distinguished based on whether they provide a complete tuning range -- $\geq\pi$ phase shift for MZIs, or $\geq$FSR for MRRs.
    We will refer to them as strong vs. weak effects. On mainstream silicon photonics platforms, thermal tuning is the only strong effect; depletion-mode tuning with a lateral diode junction is a weak effect, not able to cover a complete tuning range.
    Barium titanate (BTO) is another promising candidate for ultralow-power~\cite{Eltes:19} phase tuning despite the exotic processes needed to integrate its crystalline form with silicon photonics. BTO tuners are typically longer than thermal tuners, increasing propagation loss per neuron; however, recent work has shown that a 220~$\mu$m shifter could be possible~\cite{Abel:19} ($V_{\pi}L / \Delta V = 4.5 / 20 = 0.22$~mm).
    Strong, low-power tuning can also be achieved with MEMS~\cite{ErrandoHerranz:20}, waveguide structures that are suspended in air. MEMS phase shifters are released by a wet underetch. Although they are short and therefore low-loss, their drive mechanisms take up significant area that cannot be used for waveguide and metal routing. MEMS mechanical responses are faster than thermal at around 100~kHz--10~MHz~\cite{ErrandoHerranz:20}.

    To illustrate the calculation of weak tuning threshold, we can consider variability values obtained by Alipour et al.~\cite{Alipour:15} and tuning range values for a vertical junction microdisk obtained by Timurdogan et al.~\cite{Timurdogan:14}. The devices had similar geometries, fundamental modes, and FSRs, making them easier to compare and potentially compatible.
    In~\cite{Timurdogan:14}, a 1.1~V bias resulted in a 270~pm resonance shift and 0.7~$\mu$A leakage current. Given the radius of 2.4~$\mu$m, this leads to an extrapolated FSR efficiency of $K=0.13$~mW/FSR. The device survived a 680~pm shift, but efficiency degraded due to reverse leakage current.
    In~\cite{Alipour:15}, the microtoroids also had an FSR of 45~nm. Post-fabrication trimming (i.e. permanent parameter changes, applied after non-ideal devices are fabricated). Their initial resonance std. dev. of 290~pm was reduced to 25~pm. Furthermore, post-fabrication trimming removes any spatial correlation represented by $\sigma^{(1)}(r)$.
    The conclusion is that this variation reduction technique crossed a threshold; it makes this weak tuning device viable for locking. The net result would be a five orders-of-magnitude reduction in expected locking power compared to thermal tuning. An important direction for device research will be demonstrating variability reduction together with depletion modulation in resonators.


    \begin{table*}
      {\centering
      \caption{Platform values for weight configuration}
      \label{tab:weight-locking}
      \begin{tabularx}{\textwidth}{c|c|c|X}
      \toprule
      Name & Variable & Value & Description \\
      \hline
      \multirow{2}{5em}{Variation}
      & \multirow{2}{4em}{$\sigma_{[FSR]}^{(0)}$}
        & 0.050 & Standard deviation of resonance offset between MRRs spaced close together (FSR-units)~\cite{Chrostowski:14} \\
      \cline{3-4}
        & & 0.0055 & Reduced MRR variability using trimming~\cite{Alipour:15} \\
      \hline
      \multirow{2}{5em}{Covariation}
      & \multirow{2}{4em}{$\sigma_{[FSR]}^{(1)}$}
        & 0.060~mm$^{-1}$ & Distance dependence of standard deviation resonance offset (FSR-units)~\cite{Chrostowski:14} \\
      \cline{3-4}
        & & 0~mm$^{-1}$ & Reduced MRR variability using trimming~\cite{Alipour:15} \\
      \hline
      \multirow{3}{5em}{MRR tuning efficiency}
      & \multirow{3}{4em}{$K$}
        & 28~mW/FSR & Embedded N-doped heater~\cite{Jayatilleka:15opex} \\
      \cline{3-4}
        & & 2.4~mW/FSR & Trench etched~\cite{Dong:10opex, Cunningham:10} \\
      \cline{3-4}
        & & 0.13~mW/FSR & Vertical junction depletion~\cite{Timurdogan:14} \\
      \hline
      \multirow{4}{5em}{MZI tuning efficiency}
      & \multirow{4}{4em}{$P_\pi$}
        & 10~mW/$\pi$ & Baseline thermal phase shifter~\cite{Khanna:15} \\
      \cline{3-4}
        & & 1.2~mW/$\pi$ & Trench etched~\cite{Dong:10opex} \\
      \cline{3-4}
        & & 100~nW/$\pi$ & Barium titanate~\cite{Eltes:19} \\
      \cline{3-4}
        & & $<$100~nW/$\pi$ & MEMS phase shifter~\cite{Edinger:19} \\
      \bottomrule
      \end{tabularx}
      }
      \vspace{-12pt}
    \end{table*}

\section{Signal Resolution} \label{sec:resolution-limited}
  In this section, we consider the laser pump power needed to achieve a certain signal frequency, $f$, and resolution, $B$, in effective number of bits. The signal frequency means the bandwidth of the waveform modulating optical power envelopes, which encodes the values of analog variables. We extend upon analog photonic link theory from~\cite{Marpaung:09} to derive analytical power use expressions and extend it to multiple channels.
  The analysis -- besides relative intensity noise -- applies identically to multiwavelength and coherent architectures because they are both based on power-encoded signals.

  \subsection{Analog photonic links}
    Further analysis rests on an understanding of laser power, resolution, gain, and nonlinearity in a single analog photonic link (APL) consisting of a modulator connected to a detector. Every photonic processor, including $N\times N$ weight matrices, has an electrical input to a modulator, a linear subsystem, and an electrical output from a photodetector. This analysis was performed by Marpaung in Ref.~\cite{Marpaung:09}. Since it is foundational, basic APL theory is rederived in Appendix~\ref{sec:analog_photonic_links}.

    A key feature of APLs is the existence of operating regimes where different sources of noise dominate the signal to noise raio, or, more precisely, the spurious-free dynamic range (SFDR). In the low power regime, thermal noise originating in the photodetector dominates because the received signal is weak. At the highest powers, relative intensity noise (RIN) from the laser dominates. These regimes and the net SFDR are plotted in Fig.~\ref{fig:sfdr_withAPD}, derived in the appendix. Whereas Marpaung sought to maximize SFDR performance in a single-channel link, here, we are interested in energy efficiency of a multi-channel link. Subsequently, we extend APL theory to multi-channel architectures.

    In digital systems, each added bit increases resolution in proportion, so the required energies are a polynomial function of bit resolution. On the other hand, the power needed to generate an analog signal scales exponentially with its bit resolution~\cite{Bankman:15}. In other words, the endeavor for energy efficiency rapidly becomes futile around 6 bits and then strictly non-viable around 10 bits. Striking this balance, we focus on the 2-6 bit regime of APLs. For reference, the TrueNorth neuromorphic electronic processor uses 4-bit weights~\cite{Akopyan:15}.

  \subsection{Single channel regimes}
    Appendix~\ref{sec:analog_photonic_links} redervives SFDR as a function of optical pump power subject to three sources of noise: thermal, shot, and relative intensity. Each noise component are critial to consider because they scale differently and therefore dominate in different operating regimes. Here, we extend this analysis to arrive at coefficients relating pump power to frequency and effective bits. Typical values of these coefficients are found in Table~\ref{tab:resolution-coefficients}.

    Analog signal resolution is stated in terms of spurious-free dynamic range (SFDR), which, roughly speaking, is the ratio of maximum to minimum resolvable signals. Unlike for digital signals, analog resolution does not depend on the number of wires or serial bit slots; however, analog signal resolution can be stated in terms of an effective number of bits corresponding to an equivalent digital signal.
    The conversion of SFDR spectral density to bits is (from Eq.~\eqref{eq:sfdr-to-bits} and Eq.~\eqref{eq:sfdr-spectral-density})

    \beq
      B \mbox{[bits]} &=& \frac{1}{10 \log 2}\frac{SFDR \mbox{[dB]} - 10\log(3/2)}{2} \\
      &=& \frac{SFDR - 1.76}{6.02} \label{eq:sfdr-to-bits}
    \eeq
    where variables are defined in Appendix~\ref{sec:analog_photonic_links}. The first term converts base 10 to base 2, and the factor of 2 comes from the fact that SFDR is an electrical power and resolution is measured in terms of voltage. The 1.76 arises due to fundamental quantization error. The SFDR from system parameters is derivid the Appendix. Here, we convert SFDR to effective bits and connect it to link power.

    \subsubsection*{Thermal regime}
      Thermal noise is due to the random motions of electrons in the receiver circuitry. From Eq.~\eqref{eq:sfdr_thermal}, the SFDR in the thermal regime is

      \beq
        SFDR \mbox{[dB Hz$^{2/3}$]} = \frac{2}{3} \Bigl[&&20\log P_{1\text{pump}} \ \ldots \nonumber \\
        && {}+10\log \left(\frac{R_b}{k_{b}T}\frac{\eta_{net}^2 M^2 R_{PD}^2}{4}\right)\Bigr]
      \eeq
      where variables are defined in the Appendix. This equation represents a ratio of signal to noise per unit of spectrum. Combining this thermal SFDR equation with the effective bits equation, Eq.~\eqref{eq:sfdr-to-bits}, results in an expression of pump power needed for a given bit value

      \beq
        10\log P_{1\text{pump}} &=& \frac{30}{2} \log2 \cdot B + \frac{30}{4}\log\left(3/2\right) + \frac{10\log f}{2} \ \ldots \nonumber \\
        && {} - \frac{10}{2}\log \left(\frac{R_b}{k_{b}T}\frac{\eta_{net}^2 M^2 R_{PD}^2}{4}\right)
      \eeq
      In linear units, this one-channel power expression is

      \beq
        P_{1\text{pump}}(B, f) &=& \sqrt{f} \cdot \frac{J^*(B, R_b)}{\eta_{net}} \label{eq:lin-laser-resolution-old}\\
        \text{where} \ \ J^*(B, R_b) &\equiv& 2^{\frac{3}{2}B} \left(\frac{3}{2}\right)^{\frac{3}{4}} \sqrt{\frac{4 k_{b}T}{R_b}}\frac{1}{M R_{PD}}
      \eeq
      where we have introduced a new term, $J^*(B, R_b)$, that links power, frequency, and resolution in the thermal noise (a.k.a. Johnson-Nyquist noise) regime for a particular receiver impedance.
      The link loss, $\eta_{net}$, is separated because it will later become a function of network size. $J^*$ has units of energy-per-root-frequency.

      The impedance, $R_b$, is an argument because it is a free design parameter. It can be designed to take on a wide range of resistance values, although its value is fixed at fabrication for a particular chip.
      When a network is meant to operate at a particular bandwidth, $f$, there is an optimal design of the junction impedance such that it allows no more than the required signal bandwidth: $R_b = (2\pi f C_{pd})^{-1}$. The capacitance is not a free design parameter, rather a circuit parasitic determined by the layer thicknesses and device sizes available on a particular fabrication platform. This means there is an additional relation for the optimal design:

      \beq
        E_{thrm}(B) &\equiv& \left.J^*(B, R_b)\right|_{R_b = (2\pi f C_{pd})^{-1}} \\
        &=& 2^{\frac{3}{2}B} \left(\frac{3}{2}\right)^{\frac{3}{4}} \sqrt{8\pi k_{b}T}\frac{\sqrt{C_{pd}}}{M R_{PD}} \label{eq:J_definition}
      \eeq
      where we have introduced a new term, $E_{thrm}(B)$, describing the laser pump power needed to support an APL of a given frequency and resolution, supposing an optimal receiver design. Like $J^*$, this term links power, frequency, and resolution in the thermal regime; unlike, $J^*$, it has units of energy -- hence our choice of the variable $E$ -- resulting in an intuitive power relation:
      \beq
        P_{1\text{pump}}(B, f) &=& f \cdot \frac{E_{thrm}(B)}{\eta_{net}} \label{eq:lin-laser-resolution} \\
      \eeq
      where $P_{1\text{pump}}$ is the pump laser power for a single APL, and $\eta_{net}$ is the transmission efficiency of the APL.

      This equation has several notable features that will carry through to the multi-channel system. Firstly, power scales exponentially with number of bits, which is characteristic of analog signaling. Strikingly, the power-resolution scaling rate of $15\log2=4.5$~dB/bit is less than that of any analog electrical link: $20\log2=6.0$~dB/bit. This difference is explained by the fact that output electrical signal power is the square of the received photocurrent and thus optical pump power. The quadratic relation between signal power and supply power also explains the square-root dependence on bandwidth in Eq.~\eqref{eq:lin-laser-resolution-old}. For a fixed receiver resistance (Eq.~\eqref{eq:lin-laser-resolution-old}), the photonic system transmits more information per Joule as its bandwidth increases; however, for a resistance that varies optimally with operating bandwidth (Eq.~\eqref{eq:lin-laser-resolution}), the information per Joule does not vary.
      Finally, the APD gain, $M$, plays a prominent role. We will discuss APDs as a key technology below.

    \subsubsection*{Shot noise regime}
      Shot noise is due to the randomness in the detection times of quantized photons. From Eq.~\eqref{eq:sfdr_shot}, the SFDR in the shot noise regime is

      \beq
        SFDR \mbox{[dB Hz$^{2/3}$]} = \frac{2}{3} \Bigl[&&10\log P_{1\text{pump}} \ \ldots \nonumber \\
        && {}+10\log \left(\frac{\eta_{net} R_{PD}}{q F_A}\right)\Bigr]
      \eeq
      Combining again with Eq.~\eqref{eq:sfdr-to-bits}, we arrive at power needed for a given bit value,

      \beq
        10\log P_{1\text{pump}} &=& 30 \log2 \cdot B + \frac{30}{2}\log\left(3/2\right) + 10 \log f \ \ldots \nonumber \\
        && {} - 10 \log \left(\frac{\eta_{net} R_{PD}}{q F_A}\right) \label{eq:shot_limited_power_log}
      \eeq
      In linear units, the equation is

      \beq
        P_{1\text{pump}}(B, f) &=& f \cdot \frac{E_{shot}(B)}{\eta_{net}} \\
        E_{shot}(B) &\equiv& 2^{3 B} \left(\frac{3}{2}\right)^{\frac{3}{2}} \frac{q F_A}{R_{PD}} \label{eq:S_definition}
      \eeq
      where we have introduced a new term, $E_{shot}$\footnote{$E_{shot}(B)$ is always a function of bits, but we will sometimes drop the argument for brevity, referring to it as $E_{shot}$. The same goes for $J^*$, $E_{thrm}$, and $F_{RIN}$.}, that links power, frequency, and resolution in the shot noise regime. All of these terms have a physical limit since $F_A$ is strictly greater than one, $\eta_{net}$ is strictly less than one, and $R_{PD}$ is strictly less than $hc/(\lambda q)$, which is 1.26~A/W at 1550~nm.

      Like in the thermal regime, received signal power increases with optical pump power squared, but, now, the noise component also increases with optical power. The result is a strong resolution scaling of $30\log2=9.0$~dB/bit instead of $20\log2=6.0$~dB/bit in analog electronics.
      Another notable feature of the expression for $E_{shot}$ is that APD gain does not appear explicitly. The excess noise, $F_A$, increases with $M$ meaning that APDs strictly increase the power needed to achieve a given resolution in the shot noise regime.

    \subsubsection*{RIN regime}
      RIN is due to random changes in the power output from carrier lasers. For the relative intensity noise (RIN) relation, we combine Eq.~\eqref{eq:sfdr-to-bits} and Eq.~\eqref{eq:sfdr_rin}:

      \beq
        &&20\log2 \cdot B + 10\log \left(3/2\right) \ \ldots \nonumber \\
        && {} = \frac{2}{3} \left[- RIN - 10\log F_A + 10\log 4 - 10 \log f\right]
      \eeq
      There is no power in this expression, so we rearrange in terms of frequency. This is the maximum frequency that can be obtained at a given bit value.

      \beq
        10 \log f \leq &&- 30 \log2 \cdot B - \frac{30}{2}\log\left(3/2\right) \ \ldots \nonumber \\
        && {} - RIN - 10\log F_A + 10\log 4
      \eeq
      In linear units, it is

      \beq
        f &\leq& F_{\text{RIN}}(B) \\
        F_{\text{RIN}}(B) &\equiv& 2^{-3B} \left(\frac{2}{3}\right)^{\frac{3}{2}}\frac{4}{F_A} 10^{\frac{-RIN}{10}} \label{eq:rin-limit-definition}
      \eeq
      where we have defined a term $F_{\text{RIN}}(B)$, the maximum viable bandwidth of an APL of a given resolution. Using typical values ($M=1$, $F_A=1$, and $RIN= -155$~dB/Hz), that maximum bandwidth is approximately: $F_{\text{RIN}}(B) \approx 2^{-3B} \ \ 6.9 \times 10^{15} \ \text{Hz}$.

      RIN imposes a hard limit on the ability of laser light to represent analog signals. This limit applies regardless of how signals are generated or detected, or how powerful the laser is.
      The resolution limit is 7.5 effective bits at 1~GHz and 5.3 bits at 100~GHz. Stated as a bandwidth limit: at 4 bits, the maximum frequency is 1.7~THz; at 6 bits, it is 26~GHz; at 8 bits, it is 410~MHz.


      Table~\ref{tab:resolution-coefficients} calculates typical values for the metrics describing required laser power as limited by thermal, shot, and relative intensity noise. The first row pertains to a chip that is fabricated with 50~$\Omega$ junction resistor, while the second row uses optimally-designed junction resistors whose optimum value varies depends on the operating bandwidth.
      $E_{thrm}$ and $E_{shot}$ describe situations where more power is needed to support more bandwidth. To be invariant quantities, they therefore must have units of energy, even though it is not obvious how they correspond to a physical packet of light or electricity.
      These quantities are useful because they can be compared directly to energies of detection and digitization that might be present around an APL. As an example, $E_{shot}$ (physical limit) means that, for a 1~GHz system, the APL would require a minimum 0.96~$\mu$W of optical power to support a 4-bit signal resolution.


    \begin{table*}
      {\centering
      \caption{Laser pump power metrics for single-channel analog photonic links}
      \label{tab:resolution-coefficients}
      \begin{tabularx}{\textwidth}{c|c|c|c|c|c|l}
        \toprule
        Regime & Coefficient\footnote{$R_{PD}=0.8$~A/W, $C_{pd}=35$~fF, $M=1$, $T=300$~K, $\lambda=1550$~nm} & $B =$ 2-bit & $B =$ 4-bit & $B =$ 6-bit & $B =$ 8-bit & Units \\
        \hline
        Thermal & $\left.J^*(B, R_b)\right|_{R_b=50\Omega}$ & 250$\times 10^{-3}$ & 2.0 & 16 & 130 & nW.Hz$^{-\frac{1}{2}}$\\
        \hline
        Thermal & $E_{thrm}$ & 820$\times 10^{-3}$ & 6.5 & 52 & 420 & fJ\\
        \hline
        Shot & $E_{shot}$ (typical) & 24$\times 10^{-3}$ & 1.5 & 96 & 6.2$\times 10^3$ & fJ\\
        \hline
        Shot & $E_{shot}$ (physical limit) & 15$\times 10^{-3}$ & 0.96 & 61 & 3.9$\times 10^3$ & fJ\\
        \hline
        RIN & $F_{\text{RIN}}$ & 110$\times 10^3$ & 1.7$\times 10^3$ & 26 & 0.41 & GHz\\
        \\ \bottomrule
      \end{tabularx}
      }
      \vspace{-12pt}
    \end{table*}

  \subsection{Multiple channels} \label{sec:multiple-channels}
    In photonic neural networks and vector-matrix multiplication (VMM), each input channel has a corresponding laser and modulator. Each signal must have the potential to fan-out to the different outputs, whether fan-out occurs in a broadcast splitter (multiwavelength architecture) or within a MZI mesh (coherent architecture). By energy conservation, fan-out carries an attenuation factor of $1/N$~\cite{Goodman:1985}. Fan-out in MZI architectures is revisited in more detail in Sec.~\ref{sec:network-power}.
    At the same time, analog summation means that output signal power can recover some of this fan-out attenuation, leading to an apparent fan-in gain. Fan-in gain is dependent on the signals' cross-correlation, so we must introduce a term to quantify cases of signal correlation.


    \subsubsection*{Fan-in with correlated signals}
      As a result of additive fan-in, the root-mean-squared (RMS) power of the electrical output signal depends on the values of the inputs, meaning that SFDR and resolution become signal-dependent.
      Every situation lies somewhere on the continuum between these three cases. These cases are
      (worst- or singular case): all received signals, after weighting, are zero except for one,
      (uncorrelated-case): all inputs have the same RMS and are statistically independent,
      (best- or identical case): all inputs are the same. The three special cases are illustrated in Fig.~\ref{fig:correlation-factor}.
      The effects of fan-out and fan-in can be stated as modifications to the total received photocurrent $I_{rec}$ from Eq.~\eqref{eq:i_rec}. We use~$\left. I_{rec} \right|_{N=1}$ to indicate the baseline value.

      \begin{IEEEeqnarray}{rClll}
        \mbox{RMS}\left(I_{rec}(N)\right) &=& N^{-1} &\left.I_{rec}\right|_{N=1} \quad \quad &\mbox{(singular case)} \\
        &=& N^{-1/2} \ \ &\left.I_{rec}\right|_{N=1} &\mbox{(uncorrelated case)} \\
        &=&  &\left.I_{rec}\right|_{N=1} &\mbox{(identical case)}
      \eeq
      where the output signal amplitude is a statistical value given by

      \beq
        \mbox{RMS}\left(I_{rec}\right) &\equiv& \sqrt{\left<\left(I_{rec} - \left<I_{rec}\right>\right)^2\right>}
      \eeq
      We can introduce a similarity variable, $\xcor$, to cover these cases such that $\xcor=0$ is singular, $\xcor=.5$ is uncorrelated, and $\xcor=1$ is identical\footnote{The $\xcor$ variable describes a concept similar to that described by the $\rho$ variable introduced in~\cite{Agarwal:16} and used in~\cite{Nahmias:20}. In the case referred to as ``fixed output precision, only positive inputs/weights''~\cite[Table 1]{Agarwal:16}, there is an embedded, unstated assumption that all signals must be identical, which is always a trivial computation.}.
      In general,

      \beq
        I_{rec}(N) &=& N^{\xcor - 1} \left.I_{rec}\right|_{N=1}
      \eeq
      where $\xcor$ can take on continuous values between 0 and 1, between the extreme cases shown in Fig.~\ref{fig:correlation-factor}.
      The exact value of $\xcor$ depends on the situation, specifically, on the cross-correlation of input signals and the weight matrix. We leave its general expression for further work.

      The multichannel modification to average received photocurrent has effects both on signal and noise. In all noise regimes, adding channels increases average photocurrent and therefore maximum signal amplitude. Noise amplitude can grow at the same rate or slower than the signal amplitude, depending on the type of noise.

      \begin{figure}[tb]
        \begin{center}
        \includegraphics[width=0.8\linewidth]{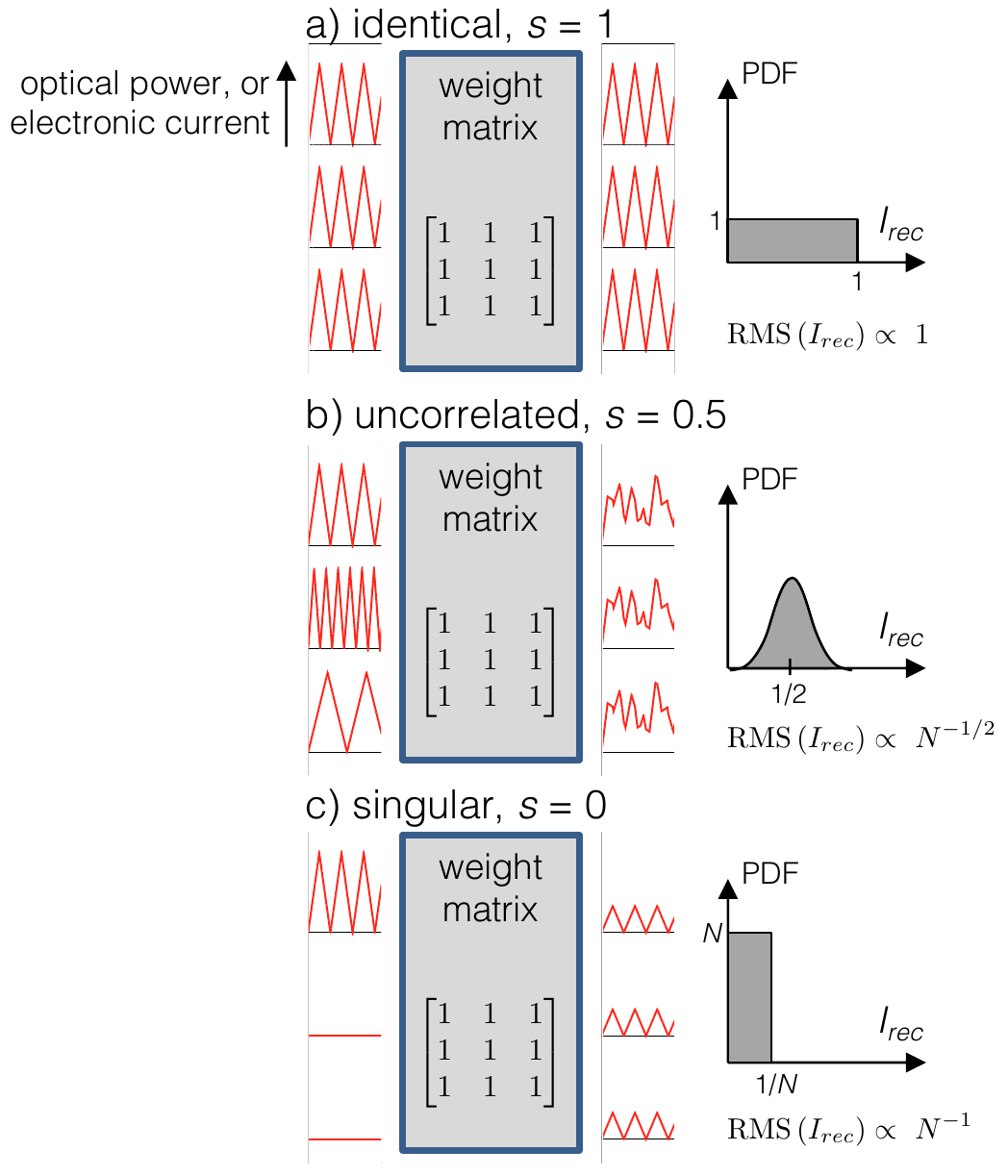}
        \caption[]{Correlation factor, fan-out, and received signal power in any physical weight matrix using power- or current-modulated signals. Left panels show different input signal cases and resulting outputs (to scale). All weight matrices are fully transmitting. Right panels show the probability density functions (PDF) of output voltage, the RMS of which determines signal power, which is compared to noise power to obtain SFDR. a) In the identical case, all power entering the matrix leaves the matrix, regardless of $N$. b) When inputs are uncorrelated, their sum approaches a Gaussian distribution whose width (i.e. signal power) decreases with $N$. c) In the worst case, only one signal is non-zero, and it must attenuate by a factor of $N$ in order to fan-out to $N$ output ports. This behavior is a consequence of photon and charge conservation and holds regardless of weight matrix implementation.}
        \label{fig:correlation-factor}
        \end{center}
      \end{figure}

    \subsubsection*{Thermal regime}
      No matter how many channels are present, there is still only one photodetector per channel, so thermal noise does not depend on $N$. Correlation-dependent fan-in only improves SFDR. We can calculate this effect by substituting the new $I_{rec}$ in Eqs.~\eqref{eq:oip_vs_irec}, \eqref{eq:full-sfdr}, and \eqref{eq:p_thrm}, and so on to reach a new version of Eq.~\eqref{eq:J_definition}.


      \beq
        P_{1\text{pump}}(N, f, B) = N^{1-\xcor} P_{1\text{pump}}(1, f, B) \quad \text{(thermal)}
      \eeq
      That means, to maintain the SFDR and thus resolution, in the worst case, the laser power must increase in proportion to fan-out, $N$. In the uncorrelated signal case, it must increase by only $\sqrt{N}$.

      In a $N\times N$ photonic network, total laser pump power carries an additional factor of $N$ to provide power to all of the input channels.
      The total laser power required by the entire network is thus

      \beq
        P_{N\text{pumps}, thrm-limit} &=& N^{(2-\xcor)} f \cdot \frac{E_{thrm}(B)}{\eta_{net}} \label{eq:resolution-limited}
      \eeq
      The result depends significantly on the signal correlation variable, $\xcor$. The result is that thermal noise-limited system power in a situation-dependent continuum somewhere between activity-proportional (best case) and MAC-proportional (worst case).

    \subsubsection*{Shot regime}
      Shot noise depends on the total received power regardless of its wavelength, so shot noise is not independent of~$N$. From Eq.~\eqref{eq:p_shot}, we see that shot noise power is proportional to $I_{rec}$, which means that it scales as

      \beq
        p_{shot}(N) &=& N^{\xcor - 1} \left.p_{shot}\right|_{N=1}
      \eeq
      Making a similar rearrangement to get needed power for a single-channel APL and an $N\times N$ network,

      \beq
        P_{1\text{pump}}(N, f, B) &=& N^{1-\xcor/2} P_{1\text{pump}}(1, f, B) \quad \text{(shot)} \\
        P_{N\text{pumps}, shot-limit} &=& N^{(2-\xcor/2)} f \cdot \frac{E_{shot}}{\eta_{net}} \label{eq:shot-limited}
      \eeq
      Shot noise dominated power is therefore MAC-proportional at worst.
      Compared to the thermal regime, the favorable effect of analog summation scales less strongly with $\xcor$.


    \subsubsection*{Coherent RIN regime}
      Relative intensity noise manifests in a fundamentally different way between multiwavelength and coherent architectures. Whereas thermal and shot noise have to do with detection, which is the same in each, relative intensity noise pertains to the optical domain where wavelength matters.

      In order to create a coherent phase state between channels, coherent weighting architectures must have exactly one laser source. When a single laser is used, the noise in the optical domain is correlated over all channels, so noise grows at the same rate of signal amplitude. Equation~\eqref{eq:p_rin} is applicable.
      As a result, the RIN-limited SFDR stays constant with $N$
      \beq
        f \leq F_{\text{RIN}}(B) \quad \text{(coherent)}
      \eeq


    \subsubsection*{Multiwavelength RIN regime}
      When multiwavelength architectures use different lasers for each channel, their intensity fluctuations are uncorrelated. The sum of uncorrelated fluctuations grows more slowly than the sum of partially correlated signals, so Eq.~\eqref{eq:p_rin} is \textit{not} applicable.
      Take, for example, the identical case of $\xcor=1$: the received photocurrent is multiplied by $N$ but RIN is multiplied by $\sqrt{N}$ because it is a sum of uncorrelated random variables.

      This RIN growth rate is the same as for shot noise, but the reasoning is different. Shot noise stems from photon detection, so it depends on the square-root of the sum of received optical signals. RIN depends on the sum of noise across received optical signals (not the square-root), but these signals each have their own independent noise components. The sum of these noise components grows only with the square-root of the number of channels.
      For an $N$-to-1 fan-in circuit as well as an $N \times N$, the RIN limit becomes

      \beq
        f \leq N^{\xcor/2} F_{\text{RIN}}(B) \quad \text{(multiwavelength)}
      \eeq

      Relative intensity noise plays an important role in capping the maximum viable frequency of an analog photonic network. This is one aspect of the differences between multiwavelength and coherent architectures. Taking, for example, a 32-channel network with 6-bit signals that are uncorrelated ($\xcor=0.5$), a single-source architecture would be capped to 26~GHz and multi-source WDM architecture to 62~GHz. It is worth noting that multiwavelength architectures can be driven by single multiwavelength lasers~\cite{Xu:20}. In these cases, the RIN over channels is correlated, and the single-source limit is applicable.

  \subsection{Foreseeable technology} \label{sec:foreseeable-APDs}

    We identify waveguide-integrated APDs~\cite{Martinez:16,Assefa:2010} and low-noise pump lasers~\cite{Morton:18} as a critical technology for neuromorphic photonics.
    The effect of APD on link parameters is shown in Fig.~\ref{fig:sfdr_withAPD}. APD gain pushes the thermal limited SFDRs to the left. In other words, they reduce the pump power needed to get a certain SFDR but do not increase the maximum achievable SFDR constrained by RIN.
    APDs amplify the signal in the photocarrier domain, between photons and photocurrent. Once the signal is a photocurrent in a circuit, thermal noise arises, after which electronic amplifiers can only degrade SFDR. APDs contain an extra dopant layer where electrons or holes are energetic enough to create more free pairs. In turn, these pairs create more free pairs and so on, hence the name avalanche. An APD comes with an additional noise factor, $F_A$, Eq.~\eqref{eq:fano-factor}, that multiplies optical noise (shot and RIN) but does not contribute to thermal noise.

    For concrete values, we will below consider the device from Martinez et al.~\cite{Martinez:16} because it requires no change to the baseline silicon photonic process. It's maximum gain was $M=60$, although its $M=10$ operating point is more favorable overall for reasons described in Sec.~\ref{sec:overall-operating-regimes}.

    Relative intensity noise in lasers imposes a hard tradeoff between bandwidth and resolution. This RIN limit could be improved using low-noise lasers. In Ref.~\cite{Morton:18}, RIN was improved from its typical --155~dB/Hz to --160~dB/Hz, which would result in a bandwidth limit increase of 3.2x at a given resolution. RIN can instead be canceled in a balanced detection architecture~\cite{Marpaung:09}, but this is a challenging prospect because it would require an exact copy of the entire neural network architecture.


\section{Gain Cascadability} \label{sec:gain-limited}
  For a computational gate or neuron to work within a larger system, each neuron must be capable of driving the next neuron, referred to as cascadability. Many nonlinear devices can be made to exhibit input-output relationships that are neuron-like, but this does not mean that device can be connected with other like devices.

  Cascadability has both a physical aspect -- the optical carrier properties must be compatible from input to output -- and a signal amplitude aspect. A cascadable photonic neuron must be able to amplify upstream signals to drive one or more downstream neurons at an equivalent strength. This is the cascadability condition, stated as $g \geq 1$, where $g$ is the differential optical-to-optical gain:

  \beq \label{eq:cascadability-condition}
    g = \left.\frac{dP_{out}}{dP_{in}}\right|_{P_{out} = P_{in}}
  \eeq

  We consider cascadability in a more definite way in reference to a particular circuit called the autapse: a neuron with one connection fed directly back to itself. An autapse based on a MRR modulator neuron is pictured in Fig.~\ref{fig:modulator_autapse}b. It is an ideal circuit for studying cascadability because its input and output are the same by definition. Therefore, it gives insight into indefinite cascadability. We studied autapse behavior experimentally in Refs.~\cite{Tait:17,Tait:18modulator} and, here, explore its relevance to power metrics and larger neural networks.

  \subsection{Autapse energy}
    The cascadability condition requires that $g \geq 1$, which leads to a minimum pump power. This pump power for modulator-class neurons was derived in Refs.~\cite{Tait:17,Tait:18modulator}, so we omit the derivation here but add in a term for APD gain, $M$.

    \beq
      \left.P_{1\text{pump}}\right|_{g=1} &=& \frac{2 V_\pi}{\pi M R_{PD} R_b} \label{eq:pump-power}
    \eeq
    where $P_{1\text{pump}}$ is the optical power entering the modulator, and $R_{PD}$ is photodiode responsivity. $V_\pi$, called the $\pi$-voltage, characterizes the modulation slope efficiency ($dT/dV$). A modulation voltage of $V_{\pi}$ is needed to induce a $\pi$ phase shift in a straight waveguide modulator. This shift corresponds to a 0-1 change in transmission in a Mach-Zehnder modulator. It is not the same as the voltage to induce a $\pi$ shift around an MRR modulator but can be used to quantify MRR modulation slope efficiency. $R_b$ is the impedance of the PD-modulator junction, which can be set externally. If no APD is used, then $M=1$.

    There is a problem in using Eq.~\eqref{eq:pump-power} as a metric because $R_b$ is a free design parameter. If the designer chooses a higher $R_b$, less pump power is needed, and bandwidth decreases. Bandwidth is determined by $f = (2 \pi R_b C_{mod})^{-1}$, where $C_{mod}$ is the total modulator capacitance. We can make an invariant metric by canceling this frequency

    \beq
      E_{aut} &\equiv& \frac{\left.P_{1\text{pump}}\right|_{g=1}}{f} \label{eq:eAut_definition} \\
      &=& \frac{4 C_{mod} V_{\pi}}{M R_{PD}} \label{eq:eAut_value}
    \eeq
    where we have called this quantity the ``autapse energy,'' $E_{aut}$. The new expression contains only device properties, so it is now invariant for a particular platform. Autapse energy does not correspond to any isolated operation; it has units of energy only because it is the ratio of a power to a frequency.

    One favorable property of autapse energy is that it can be measured with the same experiment used to demonstrate cascadability. With sufficient gain, the autapse undergoes a readily observable cusp bifurcation from monostable to bistable dynamics. This power is the cusp power, $P_{cusp}$. It was shown in Ref.~\cite{Tait:18modulator} that the conditions for cascadability and bifurcation are the same, illustrated in Fig.~\ref{fig:modulator_autapse}c. Measurement proceeds by building an autapse, increasing injected power until this qualitative transition occurs, and then measuring the power entering the system. The autapse energy is then the ratio of provided power to input signal frequency.


    \begin{figure}[tb]
      \begin{center}
      \includegraphics[width=.99\linewidth]{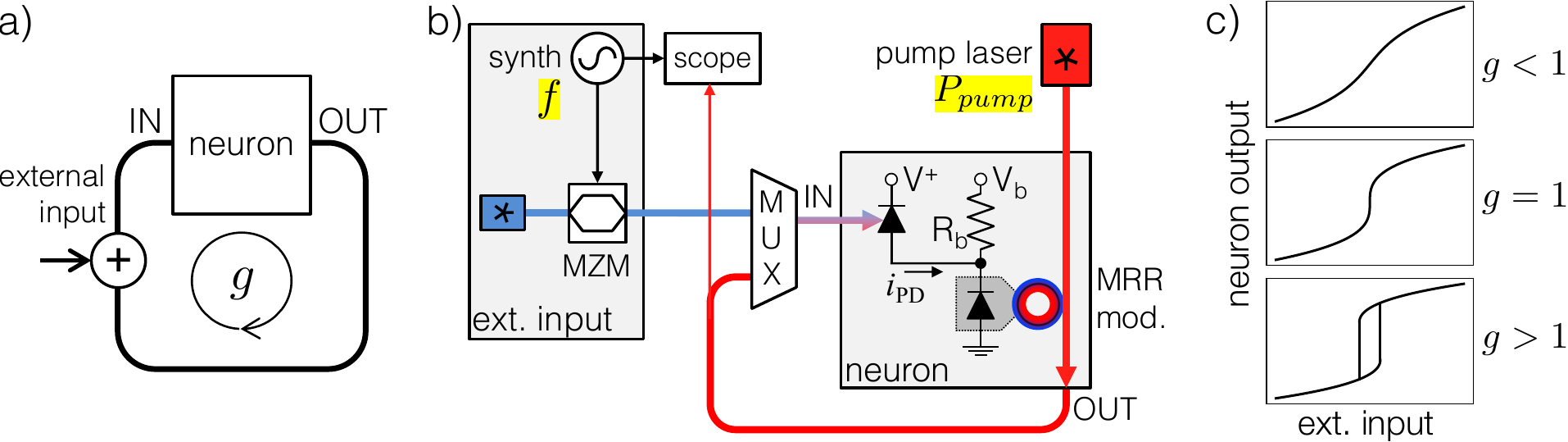}
      \caption[]{A photonic autapse used to measure a device gain and bandwidth metrics experimentally. a) An autapse is a neuron fed back to itself with a unitless round-trip gain. b) MRR modulator neuron with a feedback connection and an external input, adapted from~\cite{Tait:18modulator}. The modulator is pumped by a laser using $P_{1\text{pump}}$. The external input is a sinusoid or triangle wave with frequency $f$. c) Cusp bifurcation in response to increasing pump power. With a weak pump, the response is monostable and nonlinear. At a certain pump power, the round-trip gain crosses 1. The transition to a bistable transfer function is evidence of this crossing.
      Autapse energy is measured by this experiment as $E_{aut} = \left.P_{pump}\right|_{g=1} / f$.}
      \label{fig:modulator_autapse}
      \end{center}
    \end{figure}

    \subsubsection*{Gain in VMMs}
      VMMs consisting of similar photonic hardware with the addition of digital inputs and outputs have a  modified gain condition.
      VMMs still have a unitless electrical-in to electrical-out gain, so they yield to this gain analysis.
      Instead of being set by cascadability ($g\geq1$), the gain condition is set by the minimum voltage swing detectable on the input ADC to the maximum voltage swing provided by the input DAC.
      \beq
        g_{\text{DA-AD}} &=& \frac{V_{min, ADC}}{V_{max, DAC}}
      \eeq
      where $g_{\text{DA-AD}}$ can serve as a performance metric of the pairing of DAC and ADC technologies. VMMs are not neural networks, so they have no cascading layers or autapses.
      Despite there being no autapses, the term we called autapse energy is simply a quantification of unitless gain ($g$) of a modulator/detector platform. That means we can define

      \beq
        E_{aut, VMM} &=& g_{\text{DA-AD}} \frac{\left.P_{1pump}\right|_{g=1}}{f} \\
        &=& g_{\text{DA-AD}} \frac{4 C_{mod} V_{\pi}}{M R_{PD}}
      \eeq
      where $E_{aut, VMM}$ serves the same role as $E_{aut}$, even though there are no explicit autapses in the VMM. Using this slightly modified VMM autapse energy, all of the below scaling analysis and above noise analysis applies to VMMs without modification.

  \subsection{Cascadability in networks} \label{sec:network-power}
    The autapse is a linear chain of optoelectronic elements, but, of course, we are interested in the performance of networks of many neurons. $N \times N$ networks also have a cascadability condition, which we can state in terms of the autapse energy. Meeting the cascadability condition is essential in networks; otherwise, activity will attenuate to zero over time (recurrent topologies), to zero over layers (feedforward topologies), or below a detectability threshold (VMM systems). These conditions are always true in the worst case, although various neural network applications have been found to require less than unity cascadability within each layer~\cite{Garg:21}.

    Previous works have derived required laser power in networks optimistically, $P \propto N$~\cite{George:19,Williamson:20,Nahmias:20,Hamerly:19}, including in analog electronics~\cite{Agarwal:16}.
    We argue based on energy/charge conservation that the more pessimistic relationship, $P \propto N^2$, is the only reasonable baseline to describe any implementation of general-purpose neural networks or VMMs based on optical power- or current-modulated signals.

    \subsubsection*{All-ones matrices}
      When interconnected, the optical signal from a neuron fans out to $N$ downstream neurons. To conserve energy, this always incurs a $N$-fold optical fan-out attenuation~\cite{Goodman:1985}\footnote{A fan-out device where more energy exits than enters has been proposed~\cite{Hamerly:19} but is not considered here.}.
      Fan-out attenuation can be countered by fan-in gain. Fan-out attenuation and fan-in gain cancel completely only in one very special and computationally trivial case shown in Fig.~\ref{fig:correlation-factor}a: a maximally transparent weight matrix of all ones (``one'' being normalized against insertion loss) and identical inputs ($\xcor=1$). In this case, network energy scales with $NE_{aut}$.

      In order to be general-purpose, all neural networks and VMMs must be able to implement any weight matrix with elements from --1~to~1. This includes the all-ones matrix. They also must be able to handle any combination of input signals, including the $\xcor=0$ case, where fan-in gain is zero.
      To meet this condition, each neuron must be able to drive all downstream neurons. The cascadability condition thus becomes $g \geq N$ for every neuron, and the aggregate pump power needed to meet cascadability becomes

      \beq
        P_{N\text{pumps}, gain-limit} = N^2 f \cdot \frac{E_{aut}}{\eta_{net}} \label{eq:gain-limited}
      \eeq
      where $\eta_{net}$ is excess insertion loss of the interconnect, described in more detail below.

      Previous works have found this scaling relation to be $N$-proportional by various means. In~\cite{George:19}, this was found by imposing a cap of $1/N$ on the absolute value of all weights. We argue that this is not a useful definition because the optical power available to each receiver approaches zero as $N$ increases in all but the $s=1$ case. Even if the cascadability condition could be relaxed in this way, thermal noise (Sec.~\ref{sec:resolution-limited}) would quickly come to dominate. Reference~\cite{Williamson:20} proposed an O/E/O neuron that countered optical attenuation with electronic gain. At low gains, amplifier power consumption is constant with gain, which results in an apparent $N$-proportionality in the network power. Eventually, however, amplifier power consumption does scale with gain, which leads to that additional factor of $N$. We discuss how our analysis could extend to electronic amplifiers in Sec.~\ref{sec:further-directions}.
      Others have arrived at $N$-proportionality through a ``fixed precision'' argument~\cite{Nahmias:20}, originally outlined in~\cite{Agarwal:16}. In this argument, there is an embedded assumption that the sum of uniformly distributed random variables is also uniformly distributed, which is incorrect in all but the trivial $s=1$ case, as can be seen by comparing Figs.~\ref{fig:correlation-factor}a and b. The sum of non-identical random variables approaches a Gaussian distribution whose variance (i.e. signal amplitude) decreases with number of inputs.

    \subsubsection*{Unitary and permutation matrices}
      The quadratic scaling of gain-limited pump power applies both to multiwavelength and coherent architectures. This is readily apparent for the multiwavelength architecture because of the explicit fan-out in their broadcast splitter, but it is less apparent for MZI mesh architectures unless considering the need to handle general cases of signal vectors and weight matrices.

      MZI meshes are unitary optical devices in which the total output power can be equal to the total input power and distributed in an arbitrary way over the output waveguides. It has been argued that, therefore, all unitary matrices are also special cases for MZI VMMs which result in $N$-proportional power scaling~\cite{Shen:17}. Any matrix can be made unitary by scaling the matrix, an approach taken in Ref.~\cite{Jing:17}. This network-level weight scaling corresponds to a network-level insertion loss. In addition, unitary optical devices can be lossless only in the trivial case of identical ($\xcor=1$) inputs.

      In systems carrying power-modulated signals ($\xcor<1$), on the other hand, light entering different input ports is modulated by signals from different channels. These modulations are not identical and not interchangeable in the same way that unmodulated light is indistinguishable in unitary optical devices.
      That means there must be a fan-out of modulated energy within the mesh, even if there is not an explicit fan-out device. Every signal needs enough power to potentially reach all detectors with sufficient strength.
      It follows that, in order to cover the general case, each channel must provide $N$-proportional power, resulting in an $N^2$-proportional total laser power, the same scaling law that applies to multiwavelength architectures.

    \subsubsection*{Insertion loss}
      Insertion loss is represented by the $\eta_{net}$ variable. It includes waveguide propagation loss, which depends exponentially on waveguide propagation length.
      Propagation length is proportional to the number of neurons times either the pitch of the MRRs or length of MZIs.
      This means insertion loss is non-dominant up to some threshold in $N$, at which point, it becomes strongly dominant.
      For MRR calculations below, we assume a 20~$\mu$m MRR pitch, a 50~$\mu$m MZI length, and 1.0~dB/cm waveguide propagation loss.
      MRR weight banks also incur a fixed insertion loss due to an imperfect ability to distinguish wavelengths. As determined in Refs.~\cite{Tait:16scale,Tait:18twopole}, a typical weight bank insertion loss is 3~dB.
      Insertion loss is explicity present as $\eta_{net}$ in all relevant expressions below because waveguide loss is present in all photonic systems.

    \subsubsection*{Sub-all-to-all networks}
      Sub-all-to-all networks in which there are less than $N \times N$ possible connections will very likely be practiced to avoid quadratic hardware and power scaling. A fixed-fanout analysis was performed in Ref.~\cite{Shainline:17}. In general topologies, for one neuron to be cascadable to all of its downstream neurons $g_i \geq N_{FO, i}$, where $N_{FO, i} \leq N$ is the fan-out of neuron $i$. Summing over all neurons,

      \beq
        P_{N\text{pumps}, gain-limit} \propto \Sigma_{i=1}^N N_{FO, i} \label{eq:power-per-fanout}
      \eeq
      The sum of fan-out over all neurons is the same as the total number of weights in the network. Thus, Eq.~\eqref{eq:gain-limited} can be generalized to arbitrary topologies by simply replacing $N^2$ with the number of weights in the network.

  \subsection{Foreseeable technology}
    Equation~\eqref{eq:eAut_value} has the device-level expression for autapse energy. Device values are listed in Table~\ref{tab:photonic-cascade}. We identify critical technologies as interleaved and vertical junction modulators~\cite{Timurdogan:13,Timurdogan:14}, waveguide-integrated APDs~\cite{Martinez:16}, and graphene-based modulators~\cite{Ma:17}. The best modulators for silicon photonic neural networks could be different from those that are best for optical communication.

    For concrete values, we take as baseline a lateral depletion modulator with switching charge of about 100~fC~\cite{Khanna:15}. The interleaved junction in Ref.~\cite{Timurdogan:13} was reported to have $C_{mod}=$3.3~fF and approximate drive voltage of $V_{\pi}\approx$2.2~V, resulting in charge of 7.2~fC. The vertical junction~\cite{Timurdogan:14} has higher capacitance and lower voltage resulting in 8.5~fC.
    Switching charge can be further reduced using more advanced modulators incorporating novel materials. The graphene plasmonic slot modulator studied by Ma et al.~\cite{Ma:17} could reduce charge to to 450~aC. There is a fundamental limit on modulator switching charge set by one electron (0.16~aC), but current devices are far from this limit.

    \begin{table*}
        {\centering
        \caption{Cascadability metrics of modulators}
        \label{tab:photonic-cascade}
        \begin{tabularx}{\textwidth}{r|c|c|X}
        \toprule
        Name & Variable & Value & Description \\
        \hline
        \multirow{3}{6em}{$\pi$-voltage}
        & \multirow{3}{4em}{$V_{\pi}$}
          & 1.5~V & Voltage to induce a half-period shift in an equivalent MZM modulator. (Baseline~\cite{Khanna:15}) \\
        \cline{3-4}
          & & 0.5~V & Vertical junction microdisk modulator~\cite{Timurdogan:14} \\
        \cline{3-4}
          & & 0.95~V & Graphene push-pull modulator (proposed)~\cite{Ma:17} \\
        \hline
        \multirow{3}{6em}{Capacitance}
        & \multirow{3}{4em}{$C_{mod}$}
          & 35~fF & Depletion modulator capacitance (Baseline~\cite{Khanna:15}) \\
          \cline{3-4}
          & & 17~fF & Vertical junction microdisk modulator~\cite{Timurdogan:14} \\
          \cline{3-4}
          & & 0.27~fF & Graphene push-pull  (proposed)~\cite{Ma:17} \\
        \hline
        \multirow{3}{6em}{Autapse Energy}
        & \multirow{3}{4em}{$E_{aut}$}
          & 220~fJ & Defined in Eq.~\eqref{eq:eAut_value} (Baseline~\cite{Khanna:15}) \\
        \cline{3-4}
          & & 35~fJ & Vertical junction microdisk modulator~\cite{Timurdogan:14} \\
        \cline{3-4}
          & & 1.1~fJ & Graphene push-pull modulator (proposed)~\cite{Ma:17} \\
        \bottomrule
        \end{tabularx}
        }
        \vspace{-12pt}
      \end{table*}


\section{Optoelectronic Transduction} \label{sec:optoelectronic-switching}
  In modulators and detectors, there is power dissipated in the form of current flowing through the circuit. In all photonic information processors, there is E/O conversion at the front and O/E conversion at the back in order to interface with electronic signals. A single term can combine power from the front E/O, back O/E, and any intermediate O/E/O converters (a detector directly connected to a modulator). The analysis applies to both multiwavelength and coherent architectures on a given optoelectronics platform.
  The contribution to the network power scales with number of circuits and their bandwidth, which we refer to as activity-proportional.

  \beq
    P_{oeo} &=& N f \cdot E_{oeo} \\
    E_{oeo} &=& E_{mod} + E_{det} + E_{ADC} \label{eq:OEO_energy_definition}
  \eeq
  where $P_{oeo}$ is the network total power, and $E_{oeo}$ is a new variable. The O/E/O energy has terms associated with the modulation, detection, and analog-to-digital converters (ADCs) if present.
  We will find that modulator switching energy is always insignificant. Photodetector transduction energy becomes significant when using avalanche photodiodes but not when using PIN diodes.
  With larger networks, activity-proportional ($Nf$) contributors eventually become amortized by MAC-proportional ($N^2$f) contributors. The amortization crossover point can be up to hundreds of neurons under certain conditions, so we cannot simply neglect this O/E/O contributor.

  \subsection{Modulator charging}
    The dynamic energy needed to operate a depletion modulator is related to $C_{mod}$ and $V_{\pi}$, where $C_{mod}$ is its parasitic capacitance, and $V_{\pi}$ is the voltage swing needed to modify the transmission. We focus on modulator neurons in this analysis, pictured in Fig.~\ref{fig:conceptNetwork} and discuss laser neurons in Sec.~\ref{sec:further-directions}. Average power is activity dependent. In the worst case, the modulator has to charge every other time interval, just like in an optical communication system.

    \beq
      E_{mod} = \frac{1}{4} C_{mod} V_{\pi}^2 \label{eq:Eswitching}
    \eeq

  \subsection{Photocurrent flows} \label{sec:photocurrent}
    Power dissipated in each PD can be derived from its I-V activity. Appendix~\ref{sec:appendix-detectors} derives these currents, placing a novel constraint that the optoelectronics adhere to the cascadability condition. The result is a new metric describing the energy of detection in cascadable, analog photonic signal processors. Restating Eq.~\eqref{eq:E_det-expression-appendix},
    \beq
      E_{det} &=& 4 V_\pi C_{j} V_d \label{eq:E_det-expression}
    \eeq
    where $V_d$ is the bias voltage on each photodetector, described in more detail the Appendix. $E_{det}$ is favorable as a metric because it is invariant of operational variables and free design variables. It depends only on properties of the optoelectronic platform. Here, we discuss how detection energy can be compared to modulation energy and autapse energy, and the resulting implications for system design.

    All detectors must be biased at high enough voltage to avoid becoming forward biased when providing their maximum photocurrent.
    \beq
      V_d &>& i_{max} R_b \\
      &>& \frac{2 V_\pi}{\pi} \\
      E_{det} &>& \frac{8}{\pi} C_{j} V_\pi^2
    \eeq
    As a consequence of the cascadability condition, modulation energy is always insignificant compared to detection energy, regardless of either technology.
    \beq
      \frac{E_{det}}{E_{mod}} &>& \frac{32}{\pi} \approx 10.2
    \eeq

    Noticing that Eq.~\eqref{eq:E_det-expression} has a charge term, detection energy can also be situated in relation to autapse energy. The ratio determines which power contributor will dominate at different operating points.
    \beq
      \frac{E_{det}}{E_{aut}} &=& M R_{PD} V_d \label{eq:Eaut_Edet_ratio}
    \eeq
    For a PIN detector, $M=1$, and $V_d$ can be close to its minimum set by the forward bias constraint. With typical values, the energy ratio is approximately 1. Since laser pump power scales quadratically with $N$, electrical power associated with detection will rapidly become insignificant compared to laser power. That is not the case when APDs are used.

    APDs greatly increase both $M$ and $V_d$. The combined effect is a reduction in autapse energy and increase in detection energy.
    APD bias voltages have a dependence on their photoelectric gain and various device properties, so we must refer to example devices to get concrete values. Suppose we take an example operating point from Ref.~\cite[Fig. 2c]{Martinez:16} for a 500~nm multiplication width: $M=10$, $R_{PD}=0.8$, $V_d=16$~V. The detection to autapse energy ratio evaluates to approximately~128. What this means is that, for systems that use these APDs, detection energy will dominate when~$N<128$.

  \subsection{Analog to digital conversion}
    Digitization can play several roles in analog photonic information processors. In the linear category, all VMMs require digitization of all channels; field-programmable photonic arrays may require digitization of only a few~\cite{Perez:18fppa}; and multivariate RF photonics aim to reduce the number of ADCs~\cite{Tait:19multivariate}. Nonlinear photonic neurons based on O/E/O conversion can contain ADCs in order to tailor their transfer function~\cite{PourFard:20}, or they can can use electrooptic nonlinearities to avoid digitization altogether~\cite{Romeira:13,Tait:14,Nahmias:16}.

    Published ADC energies across the 2-6 bit range are above 1~pJ as of 2017~\cite{Murmann:17}.
    1~pJ is not a theoretical limit. It is obtained in the Supplementary Code by fitting an empirical envelope around demonstrated devices, a technique developed by Sundstrom et al.~\cite{Sundstrom:09}. In the decade since Sundstrom et al. performed this analysis, ADC efficiency has improved by about an order-of-magnitude, and it could continue to improve in the future (or outside of published literature).
    Supposing this empirical envelope of 1~pJ, ADC energy could exceed optoelectronic detection energy by two orders-of-magnitude, so it cannot be neglected. Both can be amortized in large networks, but that amortization happens much sooner in analog systems with less than complete digitization compared to photonic VMMs requiring ADCs on every channel.

  \subsection{Foreseeable technology}
    There is not much room for PIN detectors to improve to their physical limits, and O/E/O power is already non-dominant compared to contributors to laser power.
    The introduction of APDs, however, can substantially reduce laser power, at the same time that their higher bias voltage rapidly increase O/E/O power.
    If APDs are used, there is a balance to be struck between APD gain, APD bias voltage, and autapse energy, such that both optimal device design and biasing point depend on operating conditions ($N$, $f$).

\section{Summary and roadmap} \label{sec:summary}
  We have analyzed the silicon photonic neural network architectures of Fig.~\ref{fig:conceptNetwork}(a,d) when implemented on a modern-day mainstream silicon photonic foundry platform~\cite{Khanna:15}. There are four key results of this exercise: 1) an overall power equation, 2) the derivation of metric quantities describing performance, 3) the discovery of regime-like behavior in scaling behavior, and 4) a roadmap of key foreseeable technologies and their quantitative significance. The choice of roadmap technologies was constrained to devices that have been demonstrated in a research setting. The rationale for these choices is based partly in addressing limiting power components (see Sec.~\ref{sec:overall-operating-regimes}) and partly in feasibility to incorporate into silicon photonic foundry processes. In all regimes, power scales super-linearly with number of neurons, with the exception of the O/E/O regime. This finding conflicts with presumed scaling laws for neuromorphic photonics~\cite[Sec.~14.2.3]{Prucnal:17} and~\cite{Tait:17,Shen:17,Williamson:20,Hamerly:19}.

  \renewcommand{\arraystretch}{1.5}
  \begin{table*}
      {\centering\small
      \caption{New metrics for silicon photonic neural networks}
      \label{tab:power-tab}
      \begin{tabularx}{\textwidth}{c|c|c|c|X}
      \toprule
      Component & Prop. to & Coefficient & Equation & Value (Foreseeable) \\
      \hline
      Weight locking &
        $N^2$ &
        MRR: $P_{lock} = K \Omega(Nd)$; \ MZI: N/A &  
        \eqref{eq:locking-power} &
        14~mW (74~nW) \\ 
      \hline
      \multirow{2}{6em}{\centering Weight configuration}
        & \multirow{2}{4em}{\centering $N^2$}
          & MRR: $P_{conf} = \frac{K}{2\mathcal{F}}$
          & \eqref{eq:configuration-power}
          & 120~$\mu$W (230~nW) \\
        \cline{3-5}
          & & MZI: $P_{conf} = 2 P_{\pi, MZI}$
          & \eqref{eq:mzi-power}
          & 20~mW (200~nW) \\
      \hline
      Pumping: thermal limit ($B=4$) &
        $N^{(2-\xcor)} f$ &
        $E_{thrm} = 2^{\frac{3}{2} B} \left(\frac{3}{2}\right)^{\frac{3}{4}} \sqrt{8\pi k_{b}T}\frac{\sqrt{C_{pd}}}{M R_{PD}}$ &
        \eqref{eq:resolution-limited} &
        6.5~fJ (58~aJ: with APD $M=10$) \\
      \hline
      Pumping: shot limit ($B=4$) &
        $N^{(2-\xcor/2)} f$ &
        $E_{shot} = 2^{3 B} \left(\frac{3}{2}\right)^{\frac{3}{2}} \frac{q F_A}{R_{PD}}$ &
        \eqref{eq:shot-limited} &
        1.5~fJ (960~aJ: physical limit) \\
      \hline
      Pumping: gain limit &
        $N^2 f$ &
        $E_{aut} = \frac{4 V_{\pi} C_{mod}}{M R_{PD}}$ &
        \eqref{eq:gain-limited} &
        260~fJ (128~aJ: with APD $M=10$) \\
      \hline
      O/E/O &
        $N f$ &
        $E_{oeo} = \frac{1}{2} C_{mod} V_{\pi}^2 + 4 C_j V_\pi V_d + E_{adc}$ &
        \eqref{eq:OEO_energy_definition} &
        220~fJ (680~aJ) \\
      \hline
      Bandwidth: RIN limit &
        $N^{s/2}$ &
        $F_{\text{RIN}}(B) = 2^{-3B} \left(\frac{2}{3}\right)^{\frac{3}{2}}\frac{4}{F_A} 10^{\frac{-RIN}{10}}$ &
        \eqref{eq:rin-limit-definition} &
        1.7~THz (5.3~THz) \\
      \bottomrule
      \end{tabularx}
      }
    \vspace{-6pt}
    \vspace{-6pt}
  \end{table*}
  \renewcommand{\arraystretch}{1}

  \subsection{Overall power equation}
    The power components of a neural network are shown in Table~\ref{tab:power-tab}. The last column gives estimated coefficient values for a baseline foundry platform. In parentheses are the coefficient values given foreseeable technology currently in the research phase. The total power is the sum of each proportionality term (column 2) multiplied by the coefficient (column 5). Henceforth, we will refer to this sum as the overall power equation.

    Power is used by lasers, weight configuration, and electrical currents flowing in modulators and detectors.

    \beq
      P &=& N^2 \cdot P_{wei} + N \cdot P_{1\text{pump}} + N f \cdot E_{oeo} \label{eq:overall-power}
    \eeq
    The power needed for each MRR weight breaks down into static locking power and the power needed to configure the weight values: $P_{wei} = P_{lock} + P_{conf}$, which are expressed as

    \begin{IEEEeqnarray}{rCll}
      N^2 \cdot P_{wei} &=& N^2 \cdot K \Omega(Nd) + N^2 \cdot \frac{K}{2 \mathcal{F}} \ &\text{(multiwavelength)} \\ \label{eq:overall-weight}
      N^2 \cdot P_{wei} &=& N^2 \cdot 2 P_{\pi} \ &\text{(coherent)}
    \eeq
    where $K$ is tuning efficiency in mW/FSR, $\Omega$ is the expected amount of tuning per weight (in FSRs) needed to counteract fabrication variation, $d$ is MRR weight pitch, and $\mathcal{F}$ is finesse.

    The power needed for laser pumps is determined either by the cascadability threshold or the signal resolution.

    \beq
      N \cdot P_{1\text{pump}} = N^2 f \cdot \eta_{net}^{-1} \max \left[E_{aut}, \quad N^{-\xcor} E_{thrm}, \quad N^{-\xcor/2} E_{shot}\right] \quad \label{eq:overall-pumping}
    \eeq
    where $E_{aut}$ is the autapse energy, $\eta_{net}$ is network transmission efficiency -- dependent on $Nd$ -- $\xcor$ is a signal correlation factor from 0 to 1, $E_{thrm}$ is a thermal-dominated resolution coefficient defined in Eq.~\eqref{eq:J_definition}, and $E_{shot}$ is a shot-dominated resolution coefficient defined in Eq.~\eqref{eq:S_definition}.
    The last O/E/O term, is due almost entirely to current flowing in photodetectors.
    \beq
      N f \cdot E_{oeo} N f \cdot \left(E_{mod} + E_{det} + E_{adc}\right)
    \eeq

    The overall power equation exhibits an intricate, regime-like structure. The existence of regime-like behavior is one of the key results of our analysis. This can be viewed as a result of overall power being a sum of polynomials.
    Regime-like behavior means that a rich vocabulary of quantitative concepts are required to discuss power use aspects and the impacts of foreseeable technology on performance; this discussion cannot be boiled down to one concept of operation-based efficiency. We next discuss some of these aspects and must rely heavily on the terms in Eqs.~\eqref{eq:overall-power}--\eqref{eq:overall-pumping}.

  \subsection{Operating regimes} \label{sec:overall-operating-regimes}
    Over the operational domain (scale and bandwidth), different power components dominate, leading to operating regimes.
    In the above sections, we have discussed technological effects on individual components, but total power cannot be improved by focusing all efforts on any one power contributor. Here, we discuss total power, the interplay of regimes, and the thought process behind technology roadmapping.

    Figures~\ref{fig:tuning-track}--\ref{fig:graphene-track} plot system power and net MAC efficiency.
    Each panel represents a single future technology scenario, where some permutation of foreseeable technologies is incorporated.
    In the terminology, a ``region'' is a part of the domain surrounding a $(N, f)$coordinate.
    A ``regime'' is the entire sub-domain dominated by a particular component.
    A region always exists at the same spot and can fall in different regimes. So for example, we can refer to the lower part of the plots as the ``small-scale region.''

    In the plots, regimes are marked by different colors, as listed in the legend on the bottom.
    We can refer to all blue areas and the ``gain-dominated regime'' or ``gain-limited regime.''
    Regimes can move panel-to-panel.
    When moving between panels, if a regime shrinks, it means some technology has improved that contributor relative to other contributors. Next, we will discuss specific roadmaps for the multiwavelength architecture.

    \begin{figure*}[tb]
      \begin{center}
      \includegraphics[width=.95\linewidth]{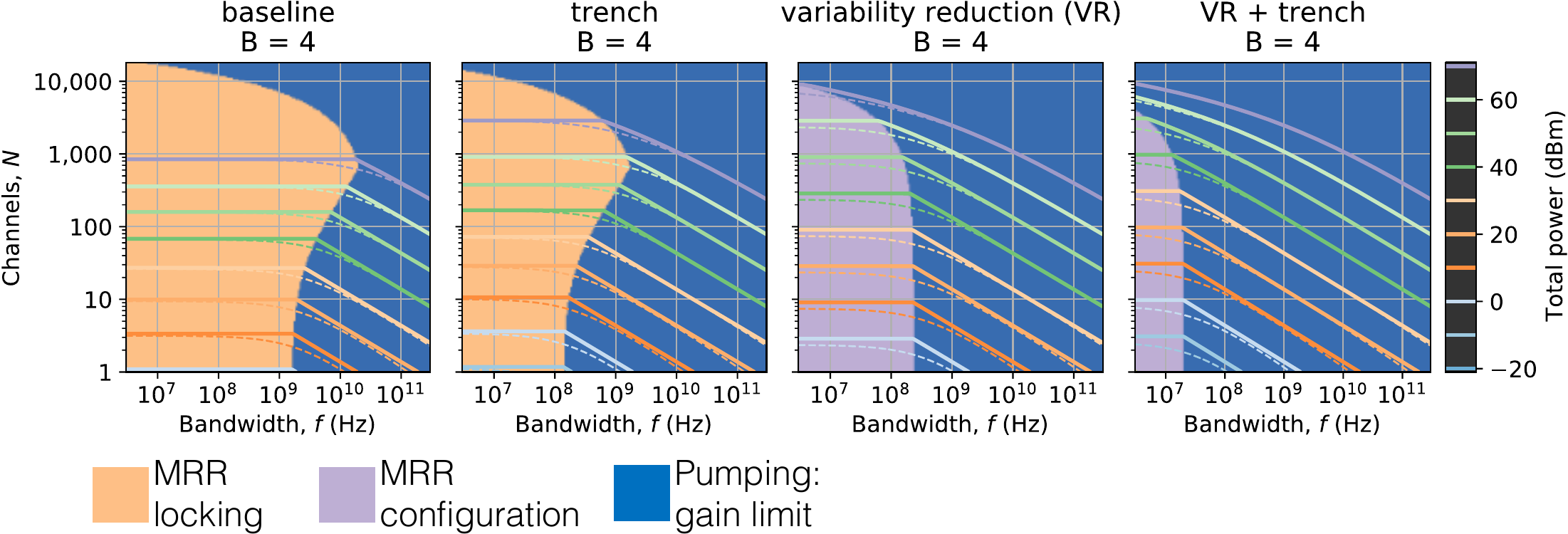}
      \caption[]{Possible roadmap for MRR tuning. Colors represent dominant power contributors. Solid contours represent power due to the dominant component. Dotted contours represent the sum of all components. a) Current day baseline for a multiwavelength architecture. b) More tuning efficiency has an effect, but locking (orange) still dominates. c) Reducing MRR variability greatly reduces locking power, but then configuration (purple) takes over. d) The impacts of both technologies compound, giving a tuning power improvement of 3 orders; however, tuning-dominated regimes also shrink. They have no impact on required laser pump power (blue).}
      \label{fig:tuning-track}
      \end{center}
    \end{figure*}

  \subsection{Technology roadmap}
    The baseline state of current silicon photonic technology~\cite{Khanna:15} is shown in Fig.~\ref{fig:tuning-track}a. There is a curved boundary between weight locking and gain pumping going through $(N, f)$ = (1, 2~GHz) and (800, 20~GHz). This curvature is due to the distance-dependence of MRR variability in $\Omega(Nd)$. Failing to account for covariation in small networks faster than 1~GHz results in an order-of-magnitude power overestimate. There is another curved boundary appearing above $N=800$. This curvature is due to a ballooning waveguide loss, $\eta_{net}(Nd)$, in networks of large physical size that must be countered by more pump power. Currently, weight locking power dominates a sizable region.

    The rest of Fig.~\ref{fig:tuning-track} shows the impacts of tuning-related technology. Trench-etched thermal tuning (Fig.~\ref{fig:tuning-track}b) moves the orange-blue boundary to the left by one order and power contours in the orange regime up by one order. When introducing a form of MRR variability reduction, such as trimming (Fig.~\ref{fig:tuning-track}c), weight locking becomes non-dominant and weight tuning (purple) is dominant. In small-scale, low-bandwidth regions, trimming reduces control power by one order; in larger-scale, low-bandwidth regions, it reduces power by two orders.
    Combining trimming and trenches (Fig.~\ref{fig:tuning-track}d) compounds the effects.

    Over-focusing on tuning-related technologies has shrunk the weight-dominated regimes to almost nothing. Over most of the domain, gain-limited laser pump power (blue) is now dominant.
    In other words, we can say that the combination of two tuning technologies has created an imbalance of regimes, where the improvements manifest only in small pockets. Next, we address that large, blue, now-limiting regime with modulator technologies.

    Fig.~\ref{fig:doping-track} shows several effects of optoelectronics affecting gain and noise: interleaved junction modulators (IJMs) and APDs. Introducing just the modulator in Fig.~\ref{fig:doping-track}a has a significant impact on $E_{aut}$; however, it shrinks the blue regime, and MRR locking dominates nearly the whole domain. The modulator alone -- just like the tuning technologies alone -- has created an imbalance of regimes, except in the opposite direction.

    Complementary technologies can be introduced to simultaneously address multiple power contributors. Fig.~\ref{fig:doping-track}b combines variability reduction with interleaved junction modulator neurons. Compare it to just trimming in Fig.~\ref{fig:tuning-track}c.
    This combination of tuning and modulator technologies has maintained a balance between $P_{conf}$ and $E_{aut}/\eta_{net}$ regimes. As a result, efficiency improves across the whole domain.

    More than being complementary technologies, the combination of IJM + VR enables non-thermal tuning, as described in Sec.~\ref{sec:foreseeable-tuning}. Non-thermal tuning partially explains the immense improvement of~\ref{fig:doping-track}b and the disappearance of tuning-dominated regimes. At this point, the whole domain is dominated by laser pump power. This component can be improved by APDs.

    \begin{figure*}[tb]
      \begin{center}
      \includegraphics[width=.95\linewidth]{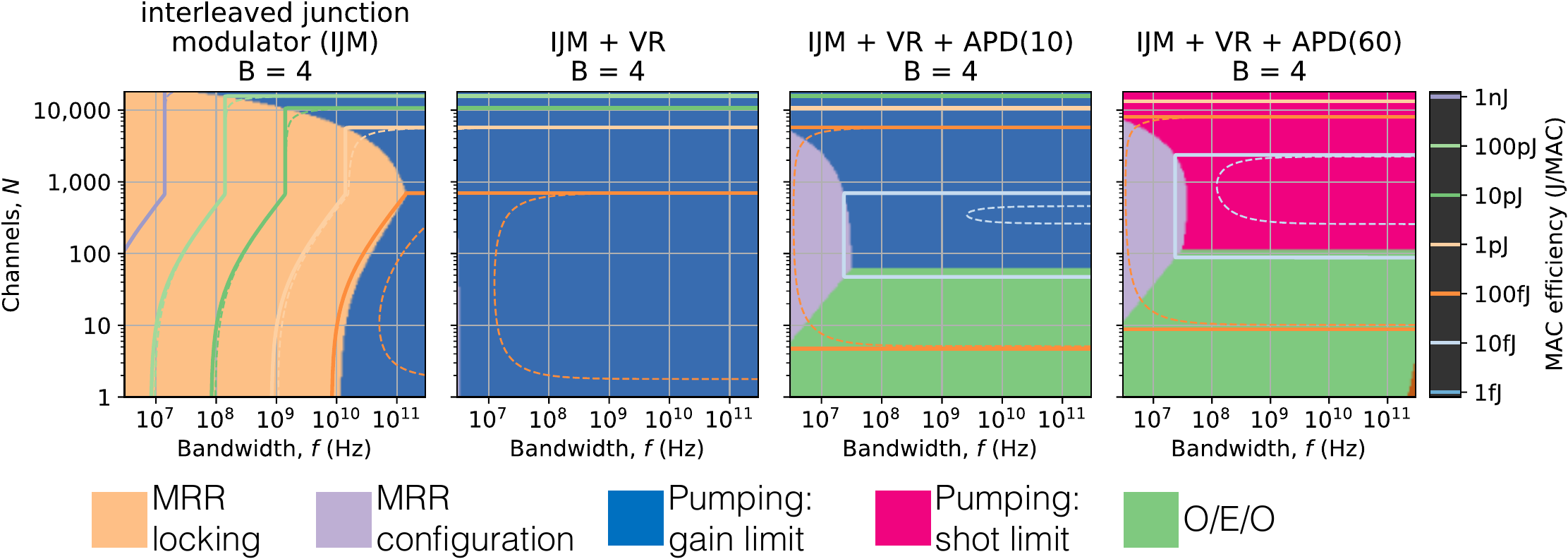}
      \caption[]{Semiconductor optoelectronics roadmap. Contours represent net MAC efficiency, the ratio of total power to $N^2f$. Solid are due to the dominant component alone, and dashed use the total power. a) Alone, interleaved junction modulators address a non-dominant regime. b) Combined with variability reduction, the modulators have a large effect both on weight tuning efficiency and the gain-dominated laser power. c) Avalanche photodetectors improve the gain limit (blue) but increase O/E/O powers (green) so that the O/E/O contributor dominates below 100 neurons. d) Increasing APD gain further brings minimal return because the shot noise limit (pink) appears.}
      \label{fig:doping-track}
      \end{center}
    \end{figure*}

    Introducing an APD reduces $E_{aut}$, so it improves the efficiency of the gain-dominated regime in Fig.~\ref{fig:doping-track}c. With $M=10$, $E_{aut}$ decreases by a factor of 10. This can be seen by comparing the horizontal, orange contour in Fig.~\ref{fig:doping-track}b to the horizontal, light blue contour in Fig.~\ref{fig:doping-track}c -- one order-of-magnitude more efficient.
    As a tradeoff, the APD increases $E_{oeo}$, so the O/E/O regime (green) now appears in the small-scale, high-bandwidth region. The horizontal green-blue boundary around $N=50$ is indicative of a constant ratio between $E_{aut}$ and $E_{oeo}$, which was derived in Eq.~\eqref{eq:Eaut_Edet_ratio}. Taking stock, we can see that these three complementary technologies have given about three orders-of-magnitude efficiency improvement across much of the domain (for example, compare the MAC contours at the $(N=100, f=10^9)$ point between Fig.~\ref{fig:doping-track}a,c).

    Fig.~\ref{fig:doping-track}d shows the effect of too much APD gain. On top of their effects on gain and detection energies, APDs affect shot noise negatively. As a result, the contours have barely moved compared to Fig.~\ref{fig:doping-track}c. What this means is that laser power must maintain at about the same level, but the reason has shifted from providing sufficient gain to providing sufficient signal-to-noise ratio. Indefinitely increasing APD gain is a dead end. In reality, APD design entails a delicate balance of three power contributors. We have included a deeper design example in the Supplementary Code.


    Fig.~\ref{fig:graphene-track} illustrates scenarios with different required resolutions. It also introduces extremely low-charge (graphene) modulators.
    The introduction of graphene into standard processes at the wafer scale poses sizable challenges; however, we consider it here because the decrease in gain-dominated power will unmask the more fundamental limits of power use in photonic neural networks. Low-charge modulators have a small impact compared to previously mentioned technologies (VR + IJM + APD). Once those technologies are introduced though, the low-charge modulators complement them, pushing efficiency further across the whole domain. Compare Fig.~\ref{fig:graphene-track}a,b to Fig.~\ref{fig:doping-track}b,c respectively. A sub-femtojoule region (deep blue contour) can be seen in Fig.~\ref{fig:graphene-track}b, although this region only exists at or below 3-bits of resolution.

    Increasing resolution slightly in Fig.~\ref{fig:graphene-track}c results in a take over by shot noise, despite the presence of all the foreseeable technologies. Fig.~\ref{fig:graphene-track}d shows a much higher 7-bit resolution. Regardless of which technologies are present, shot noise imposes a fundamental limit of 490~fJ at 7-bits, which is reflected in the contour lines of MAC energy. Furthermore, at this resolution, the RIN regime appears. RIN is not a power contributor per se, but rather a disallowed regime in terms of frequency where no amount of laser power can provide 7-bits. The RIN boundary is slanted because signals are assumed to be uncorrelated ($s=0.5$).

    \begin{figure*}[tb]
      \begin{center}
      \includegraphics[width=.95\linewidth]{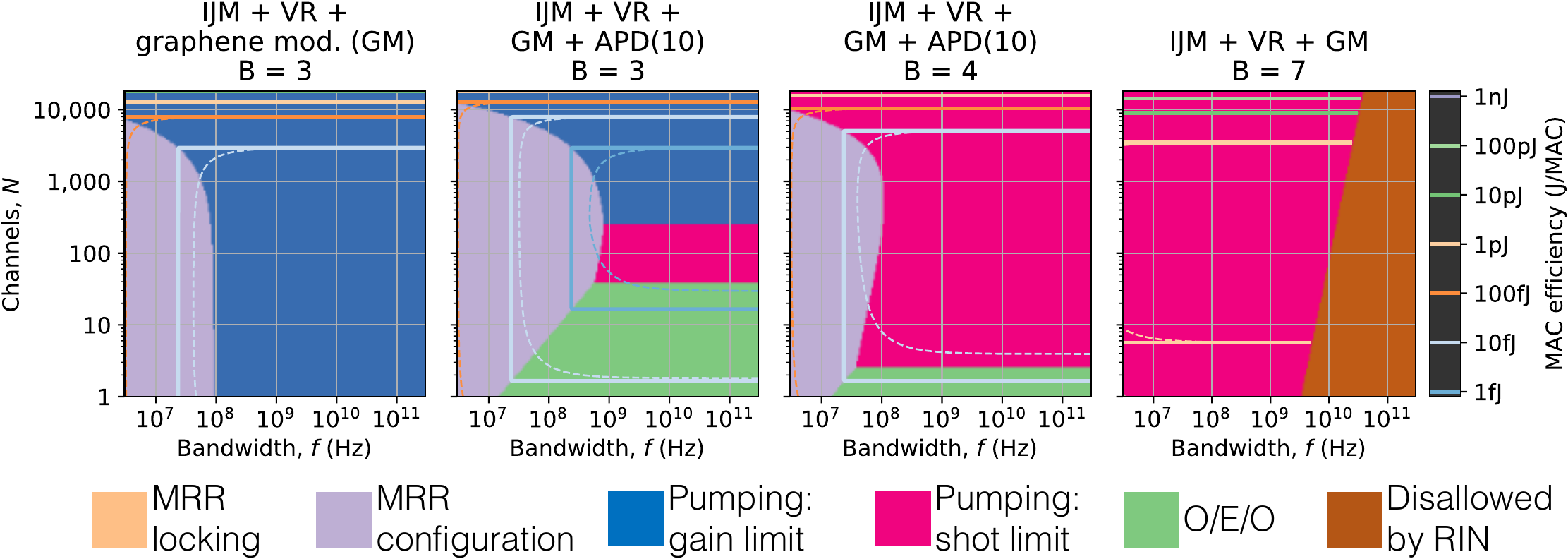}
      \caption[]{Effects of noise. a) When combined with prior technologies, low-charge graphene modulators have a substantial impact on the gain limit. b) Graphene and APDs affect different contributors and therefore compound, opening up a sub-fJ/MAC region. c) At a higher resolution, shot noise rapidly takes over most of the domain, bringing MAC efficiency above 1~fJ everywhere. d) At a much higher resolution, shot noise raises the minimum MAC energy above 1~pJ across the domain. The hard frequency limit imposed by RIN appears.}
      \label{fig:graphene-track}
      \end{center}
    \end{figure*}


  \subsection{Relevance of MACs/Joule metrics}
    Multiply-accumulate operations (MACs) and synaptic operations (SOPs) are primary metrics for evaluating electronic neural networks and VMMs.
    In systems based on electronic gates, MAC efficiency values are usually invariant with system scale and bandwidth, and this invariance makes them good metrics. In photonics, MAC efficiency can vary substantially across different regimes of operation.
    That means MAC efficiency can be an unstable lens for evaluating the performance merit of photonic systems in general.
    At a given scale and bandwidth, net operational efficiency can be obtained by dividing total power by $N^2 f$.
    For example, consider the green regime in Fig.~\ref{fig:doping-track}c. In this activity-proportional O/E/O regime,

    \beq
      P_{sys} &\approx& Nf E_{oeo} \\
      E_{MAC, sys} &\approx& \frac{E_{oeo}}{N} \label{eq:bad_MAC}
    \eeq
    This MAC/Joule expression is a poor metric for describing technology merit because it varies strongly with system scale.
    Using Eq.~\ref{eq:bad_MAC}, one might think that $E_{MAC}$ could be made arbitrarily small by imagining an arbitrarily large $N$. This type of extrapolation does not work because that imagined system would eventually enter a different operating regime.

    MAC efficiency metrics do make sense within regimes in which power scales with $N^2f$, in other words, in proportion to the rate of operations being performed. These include the gain-dominated (blue) and shot noise-dominated (pink) regimes. Visually, in Figs.~\ref{fig:doping-track}--\ref{fig:graphene-track}, MAC invariance means that contour lines are spaced far apart in those regimes. The MAC invariance breaks down again for very large $N$ due to propagation loss. Given our assumed waveguide loss value, that breakdown occurs around $N\geq$5,000.

    Shot noise imposes a physical limit on MAC efficiency. It is appropriate to use the MAC metric because it is invariant in the shot noise regime. The shot noise limit stems from the quantization of photons.
    Because light is quantized, there is a minimum number of photons needed to represent a value, regardless of how that signal is generated, detected, or propagated through any transmission element. With perfect control of quantum number, there must be at least $2^B$ quanta to represent a $B$-bit signal.
    Of course, photon detection is stochastic in classical light, and that results in a stronger scaling relationship of $2^{3B}$, which is contained in the $E_{shot}$ metric. All of the variables in Eq.~\eqref{eq:shot-limited} have physical limits.
    At 2-bit signal resolution, the physical limit is 15~aJ. At 8-bits, that limit increases by five orders-of-magnitude becoming ten times the published MAC efficiency of a 3-year-old, 8-bit electronic counterpart~\cite{Jouppi:17}. This would seem to indicate that photonic VMMs cannot compete with electronic VMMs at or above 8-bits, at least not in terms of MAC efficiency -- they would still be highly competitive in terms of latency at any resolution.

  \section{Further directions} \label{sec:further-directions}
    We have attempted to be exhaustive in identifying, exploring, and putting numbers to the factors affecting power use in MRR and MZI neural networks and VMMs. In the interest of length, we have omitted some key considerations and potential technologies in addition to other metrics. The key technologies we have identified cannot be exhaustive because new technologies are arising. It is important to note that power-related technologies may be less critical that those that affect other metrics (e.g. electronic-photonic co-packaging~\cite{Bogaerts:18}). Here, we summarize these omissions and discuss how the analyses developed can be extended to other photonic architectures.

    \subsection{Power considerations omitted}
      We have not considered wall-plug power, instead focusing on fundamental sources of on-chip energy dissipation that are needed for a photonic neural network to operate.
      In mobile applications, ``wall-plug'' power is a critical consideration.
      Furthermore, we have not made a distinction between overall power consumption and on-chip heat dissipation density. This distinction is crucial for heat sensitive environments (e.g. cryostats and datacenters) because the bulk of waste heat is dissipated by fiber-coupled pump lasers located outside of the temperature sensitive region.

      To calculate wall-plug power, the efficiency of lasers, tuning sources, and biasing sources must be incorporated. In addition, there will be power use associated with calculating weight updates and dynamically reconfigure weights. Weight calculations could become significant when reconfiguration happens at high speed, but not at low speed. Finally, there will be cooling costs that increase rapidly with heat dissipation density.

      Dissipation density is tied to device footprints. Using a MRR weight pitch of 20~$\mu$m, there is room for a $N=500$ all-to-all network on a 1x1~cm die. Today, that die at 10~GHz would consume a kilowatt. In the foreseeable future, it could consume a Watt. A 12'' wafer has room for 13k neurons in an all-to-all configuration. Of course, a wafer-scale 10~GHz all-to-all network would consume an absurd 10s of kilowatts and run into time-of-flight issues, which is why sub-all-to-all topology ideas will be indispensable.

      We have neglected some prospective technologies related to power. For example, we have not considered the potential of integrated electronic amplifiers, which were analyzed in Ref.~\cite{Lima:20}. Electronic amplifiers have a gain-bandwidth-power tradeoff, so they will effectively modify $E_{aut}$, with a penalty, without significantly changing the form of the results. Optical amplifiers do not have the same tradeoff, so they would be harder to incorporate in this analysis. Similarly, replacing modulator neurons with laser neurons would have a profound impact on autapse performance, perhaps resulting in an autapse power, rather than an autapse energy.
      Waveguides with substantially lower loss, such as silicon nitride waveguides, would directly impact the scale of feasible networks, pushing up the contours at the top edge of each plot.

      Prospective technologies for efficient weight configuration should be top priority. We have chosen to focus on post-fabrication trimming whose combination yields a four orders-of-magnitude reduction in MRR weight locking power. Other options for resonance variability reduction exist, including fabrication process improvements and robust silicon photonic resonator designs~\cite{Mikkelsen:14}.
      Other options for non-thermal tuning exist, include MEMS~\cite{Edinger:19,ErrandoHerranz:20}, Pockels effect devices~\cite{Abel:19}, and non-volatile optical materials~\cite{Cheng:17}.

      Technologies for decreasing autapse energy (i.e. neuron gain) will also play a central role.
      We have considered low-charge modulators and then APDs. APDs are favorable because they require no fabrication process modification~\cite{Martinez:16}.
      The downside of APDs is that they increase shot noise, which imposes a physical limit on MAC energy.
      In the face of the shot noise limit, it would make sense to use purely electronic transimpedance amplifiers~\cite{Lima:20,Williamson:20}.
      We did not consider electronic amplifiers, although we postulate that our analysis can be extended if an electronic platform contains some invariant gain-bandwidth-energy tradeoff. If such an invariant quantity exists, then it could be folded into autapse energy without affecting other parts of the scaling analysis.



    \subsection{Extension to other photonic architectures}
      We have developed novel power analyses for photonic neural networks and VMMs. Some of these would map to other multi-channel photonic processors for which these analyses are currently absent.

      Integrated neuromorphic photonic architectures have taken forms besides the MRR and MZI mesh approaches in Fig.~\ref{fig:conceptNetwork}.
      Deep, layered networks yield to our analytical approach because they can be analyzed as concatenated all-to-all networks, even when the number of inputs is different from the number of outputs of each layer.
      Bangari et al.~\cite{Bangari:20} proposed a hybridization of photonic and electronic fan-in, which would require more detailed analysis of the receiver junction. The WDM architectures from Feldmann et al.~\cite{Feldmann:19} and the authors~\cite{Tait:14} use a broadcast loop or bus layout, as opposed to the broadcast tree from~\cite{Tait:17}, which is used here. Bus and loop architectures put all weight banks on a single waveguide, so the waveguide propagation loss and $\eta_{net}$ would scale with $N^2$ instead of $N$.

      Sub-all-to-all topologies are subsets of the all-to-all topology and would yield to the above analyses. Sparse topologies, on the other hand, may break some of its assumptions. For example, in the brain and in the photonic neural networks considered by Shainline et al.~\cite{Shainline:18V}, average fan-out tends to 10$^4$ despite there being billions of neurons. Interconnectivity eventually scales with $1/N$, meaning a quadratic power-to-neurons proportionality will break down. Even into the future, we hazard to guess that some kind of power-to-weight proportionality, Eq.~\eqref{eq:power-per-fanout}, will hold for sparsely connected networks.

      A variety of non-neuromorphic, multi-channel photonic processing architectures have been demonstrated. Every photonic processor has input modulators and output detectors meaning the transduction analysis in Sec.~\ref{sec:optoelectronic-switching} is applicable. The analysis of tuning power in Sec.~\ref{sec:weight-control} is applicable to programmable photonics in general~\cite{Bogaerts:20}, including recirculating meshes~\cite{Perez:17,Perez:18fppa}, unscramblers~\cite{Annoni:17}, quantum optical neural networks~\cite{Steinbrecher:18}, and others. The analysis of resolution during correlated signal fan-in in Secs.~\ref{sec:resolution-limited} are extensible to microwave photonics, including field-programmable photonic arrays~\cite{Perez:17,Perez:18fppa} and multivariate photonics~\cite{Tait:19multivariate,Ma:20}. We postulate that most of these systems will exhibit regime-like scaling features that could benefit from visualization strategies in Sec.~\ref{sec:overall-operating-regimes}.

    \subsection{Signal resolution for applications}
      Signal resolution has a qualitative impact on the repertoire of neural networks.
      We have not posited what signal resolution in the $\leq$~8 bit range would be useful for various applications\footnote{Note, signal resolution and weight resolution are different concepts.}.
      Reduced-precision machine learning has shown promising results, for example, Ref.~\cite{Mishra:17} showed that accuracy on the ResNet benchmark decreased by only 0.27\% when using 2-bit signals (a.k.a. activations) and 2-bit weights.
      Outside of ML, resolution has been debated by the neuromorphic electronics community.
      The IBM TrueNorth chip uses 4-bit weights and 1-bit signals~\cite{Akopyan:15}. The BrainScales HICANN-X~\cite{Schemmel:20} chip use 6-bit weights and 5-bit signals. Pfiel et al.~\cite{Pfeil:12} demonstrated usefulness of 4-bit weights in a wide range of benchmarks and postulated that additional resolution provided little return for neuromorphic processing.
      For the photonic DEAP architecture in Ref.~\cite{Bangari:20}, a sharp jump in MNIST accuracy -- from failing to optimal accuracy -- was simulated to occur between 2- and 4- bit signals.
      Recent work has shown that certain neural network applications can be performed at dynamic resolution~\cite{Garg:21} with a corresponding reduction in noise-dominated optical power.

      We have not considered spiking photonic neurons~\cite{Nahmias:13,Romeira:13}, which have substantial differences in terms of cascadability, noise, and optoelectronics.
      To further complicate scaling analysis for spiking neurons, only those based on quiescently subthreshold sources have activity dependent power dissipation~\cite{Shainline:18II} while others based on excitable dynamics are dominated by the constant power holding them in a ready-to-fire state~\cite{Prucnal:16advances,Deng:17,VanVaerenbergh:12}.
      Finally, spikes can encode information in either rate or timing. Rate codes are continuous valued, but any tasks based on temporal coding lie completely outside of the repertoire of continuous-time photonic neurons~\cite{Peng:20}. The repertoire mismatch can undermine any kind of meaningful, quantitative comparison between the two types of architecture.


\section{Conclusion}
  We have conducted a study of power use and efficiency of fully specified architectures for silicon photonic neural networks. We find that overall power use takes a regime-like structure, which means that no less than seven metrics are needed to describe the merit of these processors. Analytical expressions for these metrics are derived and then evaluted using concrete values corresponding to demonstrated technologies. This quantitative deconstruction of power use factors has the potential to guide research and expectation in the field of neuromorphic silicon photonics.

  The scaling laws governing regimes dominated by laser pump power have finite values for MAC efficiency. These $N^2f$-proportional laws conflict with some versions of current projections and are comparitively pessimistic to those projections.
  Despite what might be seen as pessimistic performance predictions, we find that photonic neural networks and VMMs can be highly competitive in terms of MAC energy compared to state-of-the-art electronics, but only under certain operating conditions. In particular, power use increases extremely quickly with signal resolution.

  Scaling analysis links present performance values to future values enabled by new technology. We present a roadmap identifying four key technologies that will have critical impacts on total power: post-fabrication trimming~\cite{Alipour:15}, vertical depletion junctions~\cite{Timurdogan:14}, waveguide-integrated avalanche photodetectors~\cite{Martinez:16}, and graphene push-pull modulators~\cite{Ma:17}. These are specific, non-speculative technologies that have been demonstrated in research settings.
  Thermal tuning is the most pressing problem today, yet we propose a path for radical improvement for MRR-based architectures. The need to provide a certain signal gain is found to be limiting in large swaths of the future operational domain, yet it has much room to improve with foreseeable modulator technologies.

  This study reveals by example that indeed entirely new concepts and metrics will be needed to discuss the power of physics-based information processors. Instead of one number: MAC efficiency,
  there are new metric concepts that must be accounted for. The high level approach of starting from a standard all-to-all network, as opposed to starting from a device or operation, is predicted to bear fruit in the quantification of physics-based neuromorphic processors in general. In these other emerging physical platforms, we anticipate a similar yielding of efficiency invariants to efficiency regimes, although the nature of these regimes will undoubtedly be entirely different.

\newpage
\setcounter{section}{0}
\renewcommand\thesection{A\arabic{section}}
\setcounter{figure}{0}
\makeatletter
\renewcommand{\thefigure}{A\@arabic\c@figure}
\makeatother
\section{Analog photonic link derivation} \label{sec:analog_photonic_links}
  In this section, we rederive the analysis of analog photonic links (APLs) following the theory presented well by Marpaung~\cite{Marpaung:09}. An APL consists of an input RF signal, a modulator, a transmission path, and a detector. An understanding of the basic APL is foundational for arriving at resolution-frequency-power-channel relations in neuromorphic and multivariate photonics.

  The goal of this analysis is to compare analog to digital resolutions. Digital operations have an integer number of bits that determines the ratio of maximum to minimum values that can be represented. Analog signals have a similar idea of the ratio of large-signal amplitude to the smallest signal that can be resolved from noise. When signals are large enough, saturating nonlinearities in the components cause corrupting distortions.

  Spurious-free dynamic range (SFDR) is one such measure combining large-signal distortion limits and small-signal noise limits. SFDR can be converted to an equivalent bit resolution according to Eq.~\ref{eq:sfdr-to-bits}.

  \subsection{APL Saturation}
    In our rederivation, in contrast to Marpaung's, we find it fruitful to start from the track of average received photocurrent, although all below results have been verified to reproduce the results of Ref.~\cite{Marpaung:09}. The average photocurrent at the detector is

    \beq
      I_{rec} = \frac{1}{2} M \eta_{net} R_{PD} P_{pump} \label{eq:i_rec}
    \eeq
    where $P_{pump}$ is laser power into the modulator. $\eta_{net}$ is the passive attenuation of the photonic link with maximum of 1. $M$ is avalanche gain in an avalanche photodiode (APD). The responsivity as defined $R_{PD}$ takes into account the device quantum efficiency only, whereas the real responsivity of an APD is $MR_{PD}$. For a regular PIN photodiode, $M=1$.

    The RF power gain of the link~\cite[Eq.~2.25]{Marpaung:09} can be restated in terms of average photocurrent.

    \beq
      g_{RF} &=& \left(\frac{\pi R_b M \eta_{net} R_{PD} P_{pump}}{4 V_{\pi}}\right)^2 \\
      &=& \left(\frac{\pi R_b I_{rec}}{2 V_{\pi}}\right)^2
    \eeq
    where $R_b$ is the impedance of the receiver (taken here to be 50~$\Omega$), and $V_\pi$ is the voltage needed to make a Mach-Zehnder modulator (MZM) go from fully transmitting to fully blocking. $V_\pi$ can characterize modulation slope efficiency in any modulator, not just MZMs.

    Dynamic range involves the maximum signal that can be represented without distortion. Distortion arises from the electrooptic modulator transfer function, which has a saturating cubic nonlinearity. It is characterized by the output intercept point of the 3\textsuperscript{rd} harmonic ($OIP3$), which is the power at which the RF power in the cubic distorion term equals the power in the linear term. From~\cite[Eq~2.82]{Marpaung:09},

    \beq
      OIP3_{lin}\mbox{[W]} = R_b I_{rec}^2 \label{eq:oip_vs_irec}
    \eeq
    This should be surprising because there is no dependence on the shape of the modulator parameters, unlike for gain. A steeper slope of the modulator increases gain but it also increases the amount of saturation.

  \subsection{APL Noise}
    There are three sources of noise: thermal, shot, and relative intensity noise (RIN). Their power in linear watts are
    \begin{IEEEeqnarray}{rClL}
      \frac{p_{thrm}}{f} &=& k_{b} T \qquad \quad & \label{eq:p_thrm} \\
      \frac{p_{shot}}{f} &=& \frac{q R_b M F_A}{2} & \left(I_{rec} + I_d\right) \label{eq:p_shot} \\
      \frac{p_{rin}}{f} &=& 10^{\frac{RIN}{10}} \frac{R_b F_A}{4} & I_{rec}^2 \label{eq:p_rin}
    \eeq
    where $RIN$ is a laser parameter with a typical value of $-155$~dB/Hz. $I_d$ is dark current, usually less than received photocurrent. $M$ is avalanche gain. Note that these are different definitions than the ones Marpaung uses by a factor of 2 -- When adding up noise contributions under an assumption of a lossy, matched receiver, half of the noise power goes into the matching impedance rather than the load. The Fano factor, $F_A$, describes noise due to avalanche gain in an APD. $F_A$ is defined

    \beq
      F_A = K_A M + (1 - K_A) (2 - M^{-1}) \label{eq:fano-factor}
    \eeq
    where $K_A$ is the carrier ionization ratio. A typical value is of $K_A$ is 0.1~\cite{Assefa:2010}, so a typical value of $F_A$ for $M=10$ is 2.7. Avalanche noise multiplies optical sources of noise (shot and RIN) but has no effect on thermal noise. This means it only makes sense to use an APD in a thermal noise regime

    Total noise is

    \beq
      p_N \mbox{[W]} &=& p_{thrm} + p_{shot} + p_{rin} \\
      P_{N} \mbox{[dBm]} &=& 10 \log{\frac{p_{N}}{10^{-3}}}
    \eeq
    where capital $P_N$ represents the same concept except in log units. All noise terms are proportional to signal bandwidth, $f$, so they are often stated in log units as dBm/Hz. $P_N$ can be stated at a particular frequency or independent of frequency, in which case it is called ``power spectral density.''

    \beq
      \underbrace{P_N (f \mbox{[Hz]})}_{\mbox{power}} \mbox{[dBm]} = \underbrace{P_N \mbox{[dBm/Hz]}}_{\substack{\text{power}\\ \text{spectral density}}} + 10 \log(f\mbox{[Hz]})
    \eeq


    Finally, the SFDR can be calculated. Combining the noise figure (NF) and SFDR definitions found in Marpaung~\cite[Eq.~2.48, Eq.~2.85]{Marpaung:09}:

    \beq
      NF &=& P_N - G + P_{thrm} \\
      SFDR &=& \frac{2}{3}\left(OIP3 - NF - G + P_{thrm}\right),
    \eeq
    we arrive at

    \beq
      SFDR &=& \frac{2}{3} \left(OIP3 - P_{N}\right) \label{eq:full-sfdr}
    \eeq
    where $OIP3$ is now in log units. Equation~\eqref{eq:full-sfdr} is plotted in Fig.~\ref{fig:sfdr_withAPD}. As desired, the concept of a ratio between highest-power (determined by modulation saturation) and lowest-power (determined by detection noise) is now apparent. The SFDR can be stated at a given frequency and also independent of frequency, just like the noise power. SFDR has a fractional term, and so too does its frequency dependence:

    \beq
      \underbrace{SFDR (f \mbox{[Hz]})}_{\mbox{power ratio}} \mbox{[dB]} = \underbrace{SFDR \mbox{[dB Hz$^{2/3}$]}}_{\substack{\text{power ratio}\\ \text{spectral density}}} - \frac{2}{3} 10 \log f \label{eq:sfdr-spectral-density}
    \eeq

  \subsection{Operating regimes}
    Noise power is a logarithm of a sum, so the largest term is over-emphasized -- this is the origin of noise ``regimes,'' operating points where other sources of noise are not significant. At high powers, RIN dominates as it scales with $I_{rec}^2$ -- the same rate as gain. Here, $P_N \approx P_{RIN}$, so

    \begin{IEEEeqnarray}{Cll}
      \label{eq:rin_limit}
      &SFDR \mbox{[dB Hz$^{2/3}$]} \ \ldots \nonumber \\
      &= \frac{2}{3}\Bigg\{\underbrace{10\log\left(I_{rec}^2 R_b\right)}_{OIP3} - \underbrace{\left[10\log\left(I_{rec}^2 R_b\right) + RIN + 10 \log \left(\frac{F_A}{4}\right)\right]}_{P_{RIN}} \Bigg\} \nonumber \\
      &= \frac{2}{3} \left[- RIN - 10 \log F_A + 10 \log 4\right] \label{eq:sfdr_rin}
    \eeq
    That's the absolute maximum SFDR for a given RIN parameter. For typical parameters, this means $SFDR \leq 107$~dB Hz$^{2/3}$, which can be seen at high powers in Fig.~\ref{fig:sfdr_withAPD}.

    In the shot noise regime, noise increases with optical power, but more slowly that in the RIN regime. SFDR increases with the square root of optical power and gain. The shot-limited SFDR expression is

    \begin{IEEEeqnarray}{Cl}
      &SFDR \mbox{[dB Hz$^{2/3}$]} \ \ldots \nonumber \\
      &= \frac{2}{3}\Bigg\{\underbrace{10\log\left(I_{rec}^2 R_b\right)}_{OIP3} - \underbrace{\left[10 \log \left(I_{rec} R_b\right) + 10\log \left(\frac{q M F_A}{2}\right)\right]}_{P_{shot}}\Bigg\} \nonumber \\
      &= \frac{2}{3} \left[10\log \left(I_{rec}\right) - 10\log \left(\frac{q M F_A}{2}\right)\right]
    \eeq
    where the electron charge, $q$, is often written evaluated to 191 in log units. Expanding $I_{rec}$, one finds that the APD gains, $M$, cancel out.
    \beq
      &SFDR \mbox{[dB Hz$^{2/3}$]} = \frac{2}{3} \Bigg[&10\log P_{pump} + 10 \log\left(\eta_{net} R_{PD}\right) \ \ldots \nonumber \\
      &&{} - 10\log \left(q F_A\right)\Bigg] \label{eq:sfdr_shot}
    \eeq
    This means an APD leaves us with only its noise term, $F_A$, and no benefit in the shot noise regime.

    \begin{figure}[tb]
      \begin{center}
      \includegraphics[width=.9\linewidth]{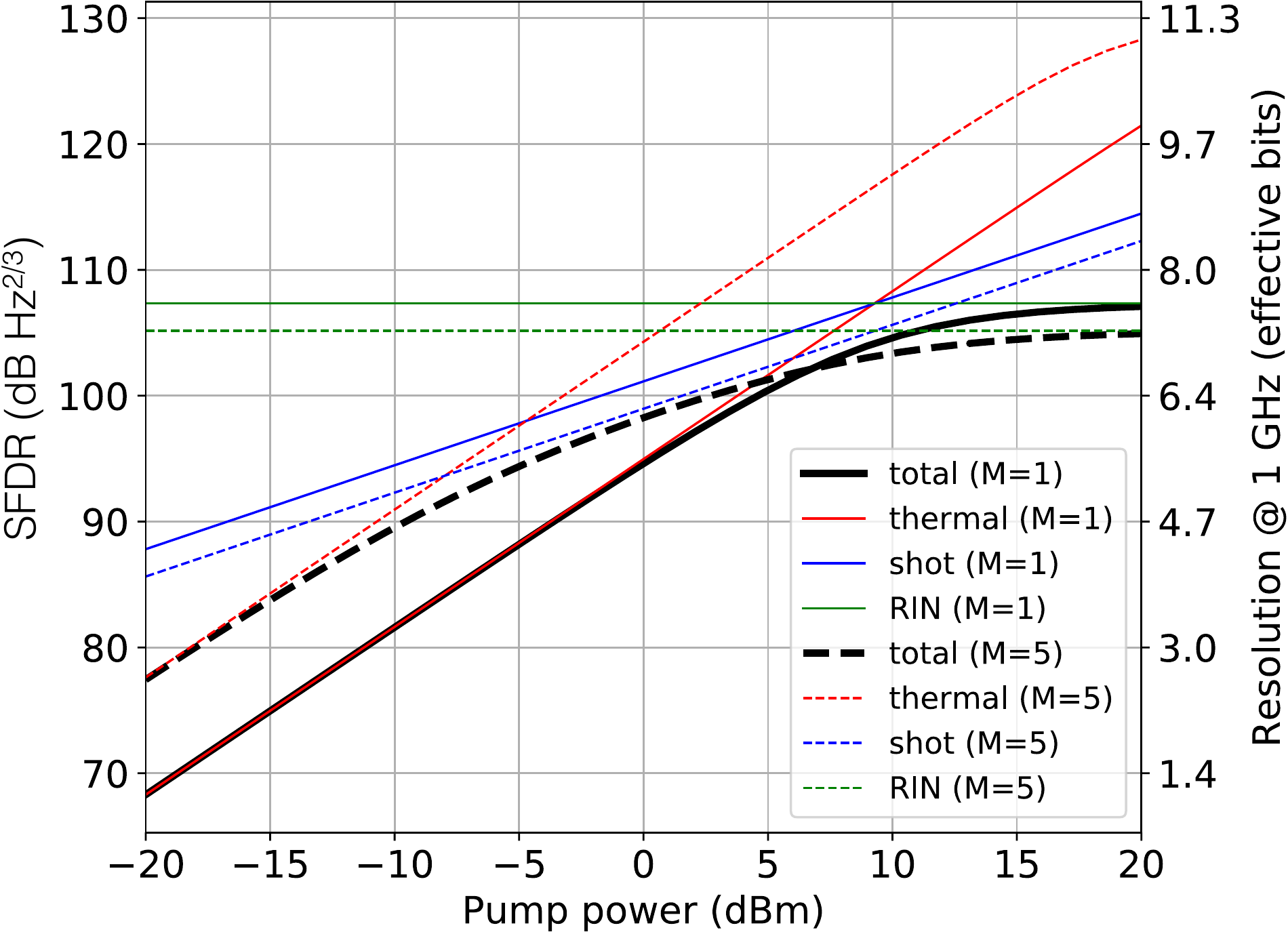}
      \caption[]{SFDR components and total for PIN receiver (solid) and APD (dashed). In the thermal regime (red), APD gain can be viewed as a leftwards shift, reducing the input power needed to achieve a given SFDR. An APD strictly degrades shot-limited (blue) and RIN-limited (green) SFDR.}
      \label{fig:sfdr_withAPD}
      \end{center}
    \end{figure}

    At low powers, thermal noise dominates. Thermal noise does not scale with $I_{rec}$, which means more gain (i.e. more received photocurrent) will proportionally increase the SFDR. Here, $P_N \approx P_{thrm}$, so

    \beq
      SFDR \mbox{[dB Hz$^{2/3}$]} &=& \frac{2}{3} \left[\underbrace{10\log \left(I_{rec}^2 R_b\right)}_{OIP3} - \underbrace{10\log \left(k_{b}T\right)}_{P_{thrm}}\right] \quad \label{eq:sfdr_thermal_outset}
    \eeq

    We replace $I_{rec}$ from Eq.~\eqref{eq:i_rec} in this low-power, thermal regime in order to state in terms of pump power.

    \begin{IEEEeqnarray}{Cl}
      &SFDR \mbox{[dB Hz$^{2/3}$]} \ \ldots \nonumber \\
      &= \frac{2}{3} \left[20\log \left(\frac{M \eta_{net} R_{PD} P_{pump}}{2}\right) - 10\log \left(\frac{k_{b} T}{R_b}\right)\right] \\
      &= \frac{2}{3} \left[20\log P_{pump} + 10\log \left(\frac{R_b}{k_{b}T}\frac{M^2 \eta_{net} R_{PD}^2}{4}\right)\right] \label{eq:sfdr_thermal}
    \eeq
    Using typical values ($R_b = 50\Omega$, $\eta_{net} = 0.32$, $R_{PD} = 0.75$~A/W, $T = 290$~K), the second term evaluates to 202~dBHz$^{2/3}$W$^{-2}$. Therefore, a pump of 1~mW with a PIN diode yields an SFDR of 94.6~dBHz, which is corroborated by the solid red line in Fig.~\ref{fig:sfdr_withAPD}.

    Analog signal resolution can be stated in terms of an effective number of bits corresponding to an equivalent digital signal, as described around Eq.~\eqref{eq:sfdr-to-bits}.
    To give a better practical sense in terms of equivalent bit values: 4-bit, 6-bit, and 8-bit correspond to SFDR of 25.8, 37.9, and 49.9dB, respectively. At 10 GHz, the frequency contribution subtracts 66.7 dB from the SFDR in dBHz$^{2/3}$. This means, for 4 bits, one would need 25.8 + 66.7 = 92.5~dBHz$^{2/3}$. For 8 bits, this is 49.9 + 66.7 = 116.6~dBHz$^{2/3}$, which is not possible, even with an APD, due to the RIN limit in Eq.~\eqref{eq:rin_limit}. The highest frequency that would work at 8 bits within the RIN limit is $10^{3/2\times(107.3-49.9)/10} = 407$~MHz.

\section{Electrical current in detector circuits} \label{sec:appendix-detectors}
  In this appendix, we derive currents that flow within resistive optical receiver circuits for photonic neural networks. The key novelty will be introducing a gain cascadability constraint. In other words, the photodetector has an output swing in terms of voltage. This voltage swing must be enough to drive a voltage-mode modulator at a sufficient strength to change optical power.

  Under cascadability constraints, an invariant energy term associated with photodetection will arise. This energy depends strongly on the type of detector: PIN vs. APD. The implication of results are discussed in Sec.~\ref{sec:photocurrent}.
  We consider both balanced detection circuits capable of positive/negative weights and single-ended receivers that allow only positive weights. They are pictured in Fig.~\ref{fig:receiver-circuits}.

  \begin{figure}[tb]
    \begin{center}
    \includegraphics[width=.4\textwidth]{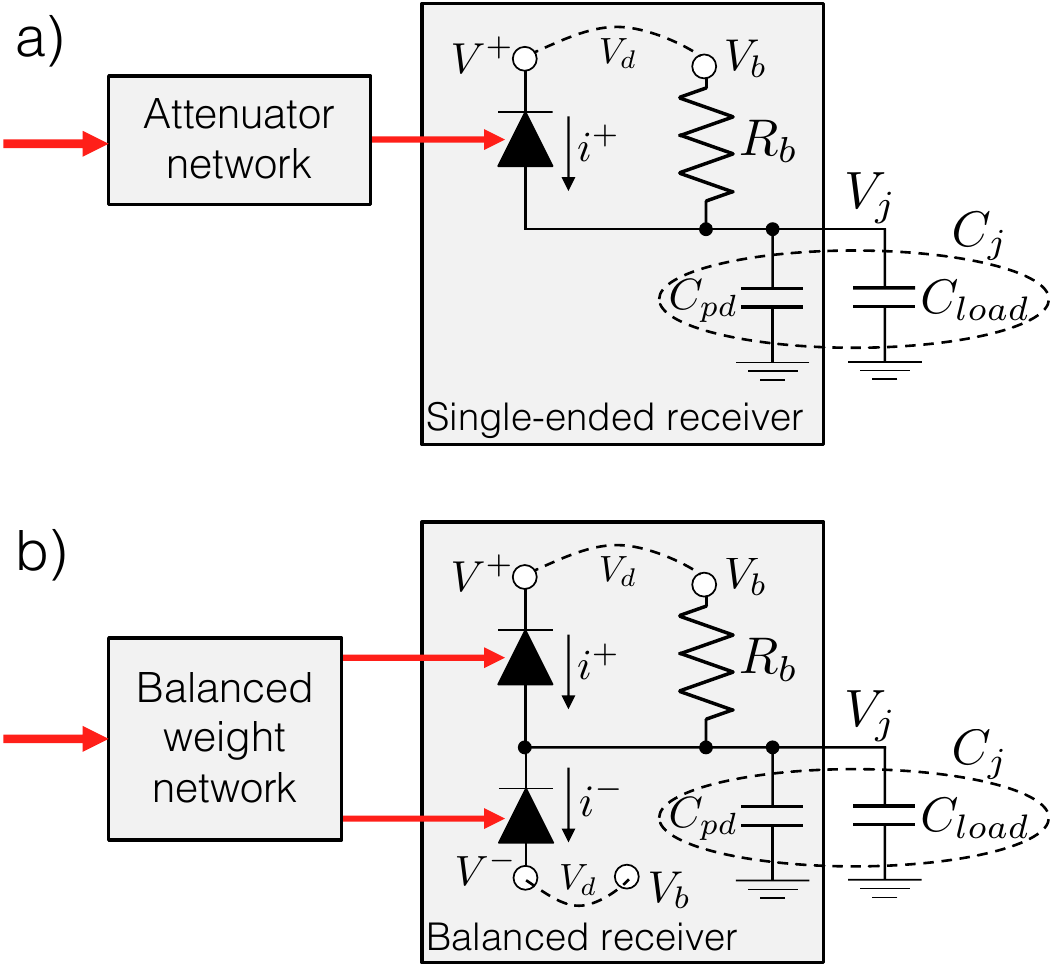}
    \caption[]{Photoreceiver circuits used to detect optical signals showing meanings of variables. Modulators are modeled as a capacitive load in parallel with PD parasitic capacitance. a) Single-ended circuit capable of positive weighting. b) Balanced receiver capable of positive and negative (i.e. signal inverting) weighting.}
    \label{fig:receiver-circuits}
    \end{center}
  \end{figure}

  Power dissipated in the balanced receiver circuit can be derived from its I-V activity.
  \beq
    P_{det} &=& \underbrace{i^+ (V^+ - V_{j})}_{\text{Positive PD}} + \underbrace{i^- (V_{j} - V^-)}_{\text{Opposing PD}} + \underbrace{(i^+ - i^-)(V_{j} - V_b)}_{\text{Bias resistor}}
  \eeq
  where variables pictured in Fig~\ref{fig:receiver-circuits}b are $P_{det}$: total circuit power, $V^{+/-}$: voltage biases on respective photodiodes, $i^{+/-}$: photocurrent in respective photodiodes, $V_j$: voltage of the common junction, and $V_b$: bias voltage on the junction. This expression simplifies to

  \beq
    P_{det} &=& i^+ (V^+ - V_b) + i^- (V_b - V^-)
  \eeq
  For convenience, we will define a variable, $V_d$ (``dark'' voltage), to represent the voltage across a photodetector when no light is present. We also suppose that PD biases are symmetric with respect to junction voltage.

  \beq
    V_d &\equiv& (V^+ - V_b) = (V_b - V^-) \\
    P_{det} &=& (i^+ + i^-) V_d \\
    P_{det, \text{single-ended}} &=& i^+ V_d \\
    P_{det, \text{balanced}} &=& i_{max} V_d \\
  \eeq
  The difference between the two circuits in Fig.~\ref{fig:receiver-circuits} is that single-ended receivers do not have an $i^-$.
  There is a maximum photocurrent, $i_{max}$, that depends on the original optical pump power. In a balanced receiver, total photocurrent is always $i_{max}$. Negative weights are obtained by changing the ratio of light directed to the complementary photodetectors. On average, the power dissipated in the single-ended architecture is half as much.
  The optical signal is generated by a pump laser, modulated, then passed through a reconfigurable transmission network, which allows us to determine maximum photocurrent

  \beq
    i_{max} = M R_{PD} \eta_{net} P_{1\text{pump}}
  \eeq
  where $P_{1\text{pump}}$ is the pump power, $R_{PD}$ is the responsivity of each PD, $\eta_{net}$ is any insertion loss of the network, and $M$ is and photoelectric gain that is present when using avalanche photodetectors (APDs).

  The cascadability condition defined in Eq.~\eqref{eq:pump-power} gives a value for pump power. Restating that equation and simplifying $i_{max}$,

  \beq
    \left.P_{1\text{pump}}\right|_{g=1} &=& \frac{2 V_\pi}{\pi M R_{PD} R_b} \\
    i_{max} &=& \frac{2 V_\pi}{\pi R_b}
  \eeq
  where $g=1$ is the cascadability condition, and $V_\pi$ is the modulation slope efficiency of the modulator that was used to generate the optical signal at the input of the transmission network. In an O/E/O neuron, the photoreceiver will also drive this type of modulator.

  Bias resistance, $R_b$, is a free design parameter, but it has an optimal design based on the intended bandwidth of operation.

  \beq
    R_b &=& \left(2 \pi C_j f\right)^{-1} \quad \text{(optimal design)} \\
    i_{max} &=& 4 V_\pi C_{j} f
  \eeq
  where $f$ is operating bandwidth, and $C_j$ is the capacitance of the circuit junction. Junction capacitance is the sum of photodetector parasitic capacitance, $C_{pd}$, and capacitance of the load, $C_{load}$. The load can be another modulator, or it could be the input of a digitizing circuit at the back end of an optical subsystem. Substituting into the above equation,

  \beq
    E_{det} &\equiv& \frac{P_{det}}{f} \\
    E_{det} &=& 4 V_\pi C_{j} V_d \label{eq:E_det-expression-appendix}
  \eeq
  where we have defined a new variable, $E_{det}$, to describe the electrical energy associated with detecting modulated optical signals within a cascadable photonic network.
  Implications of this derivation are discussed in Sec.~\ref{sec:photocurrent}.

\section*{Funding}
  This work was supported by the National Resource Council (NRC) postdoctoral fellowship program.

\section*{Acknowledgement}
  We thank Dr. Philip Yechi Ma for edits and suggestions on APL noise sections. We acknowledge the foundational role of discussions with Bhavin Shastri, Mitch Nahmias, Thomas Ferreira de Lima, and Paul Prucnal. For editorial contributions, we thank Jeff Shainline, Sonia Buckley, and Norman Sanford. This is a contribution of NIST, an agency of the US government, not subject to copyright.

\bibliographystyle{apsrev4-1}
\bibliography{powerNeuromorphic}


\newpage
\onecolumngrid
\section*{Glossary}
\renewcommand{\arraystretch}{2}
\begin{tabularx}{.8\textwidth}{rrlX}

\multicolumn{4}{c}{\textul{Physical variables}}\\
  &$k_B$ && Boltzmann constant \\
  &$q$ && Electron charge \\
  &$T$ && Temperature \\

\multicolumn{4}{c}{\textul{Operational metrics}}\\
  &$N$ && Number of photonic neurons in an all-to-all connected network; alternatively, number of neurons per layer in a fully-connected, feedforward pair of layers; alternatively, the length of the vector in a square vector-matrix computation. In all cases, the number of weights is $N^2$ at most. \\
  &$f$ && Maximum operating frequency of one neuron, one channel, or one vector element signal. Identical for input (received) and output (transmitted). Not aggregated over channels. \\
  &$B$ && Effective bits of resolution of a received signal in the electrical domain, after fan-in, i.e. one neuron input or one output of a matrix-vector multiplier. \\

\end{tabularx}
\begin{tabularx}{.8\textwidth}{rrlX}
\multicolumn{4}{c}{\textul{Weight control}}\\
  &$P_{wei}$ && System-level power consumed to configure weights and set their values \\
  &$P_{lock}$ && Static power needed to bring one microring weight onto resonance with a desired signal wavelength. Does not apply to interferometer-based optical weights \\
  &$P_{conf}$ && Configuration-dependent power needed to tune one weight over its range of values. Applies to both resonator and interferometer optical weights. \\
  &$FSR$ && Free Spectral Range. The wavelength spacing between subsequent resonances of a single optical resonator \\
  &$\mathcal{F}$ && Finesse. The sharpness of the resonance relative to the FSR. Equal to the number of similar resonators that can be cascaded without their spectra overlapping by more than 3~dB \\
  &$\sigma^{(0)}$ && Standard deviation of resonance offset between a pair of nominally identical microring resonators (MRRs) fabricated nearby to one another. \\
  &$\sigma^{(1)}$ && Coefficient of standard deviation of resonance offset as a function of distance between a pair of MRRs. \\
  &$d$ && Distance between neighboring MRRs when arranged in a square array \\
  &$\Omega(N)$ && Expected value of resonance shift per-MRR needed to bring all MRRs onto resonance with their respective signal wavelengths when they are arranged in an $N\times N$ square array \\
  &$K$ && Tuning efficiency of a MRR in terms of the power needed to tune a MRR over one FSR. Even if the MRR does not have a safe range covering an FSR, its tuning efficiency can be described by $K$ by extrapolating its wavelength shift vs. tuning power. \\
  &$P_{\pi}$ && Thermal tuning efficiency of a Mach-Zehnder interferometer in terms of power needed to redirect light entirely from one output port to the other. \\
\end{tabularx}
\begin{tabularx}{.8\textwidth}{rrlX}
\multicolumn{4}{c}{\textul{Optoelectronics}}\\
  &$R_b$ && Impedance of the photodetector. Typically determined by an external source, so it is a free parameter that can be chosen by the system designer \\
  &$C_{pd}$, $C_{mod}$, $C_j$ && Capacitance of photodetector and modulator junctions. Typically determined by the device platform. In the case of a photodetector-modulator junction, an occurence of either variable should be replaced by total junction capacitance, $C_j = C_{pd} + C_{mod}$. \textit{In this draft}, they are used interchangeably. \\
  &$R_{PD}$ && Photodiode responsivity: milliamps of current generated by one milliwatt of incident optical power. Has a theoretical maximum of 1.26~A/W at 1550~nm \\
  &$M$ && Avalanche photodiode gain, excluding quantum efficiency. The net responsivity of an avalanche photodiode is $MR_{PD}$. For a non-avalanche photodiode, $M=1$ \\
  &$F_A$ && Excess noise resulting from the avalanche process. For a non-avalanche photodiode, $F_A=1$ \\
  &$P_{oeo}$ && System-level power used for optoelectronic transduction \\
  &$E_{oeo}$ && Energy used for optoelectronic transduction for one neuron in a single $1/f$ time interval. \\
  &$E_{switching}$ && Energy consumed due to electrical-to-optical conversion in a modulator. A.k.a. energy-per-bit. In a cascadable photonic neuron, it is negligible compared to photodetection energy. \\
  &$P_{det}$ && Power consumed due to optical-to-electrical conversion in a photodetector. For a balanced photodetector (BPD), it includes the contribution of both. \\
  &$E_{det}$ && Energy consumed due do photodetection per $1/f$ time interval \\
  &$V_{mod}$ && Activity-dependent voltage of the common junction in a BPD with resistive load \\
  &$V_{b}$ && Bias voltage applied to the common junction with a source impedance of $R_b$ \\
  &$V^{(+,-)}$ && Supply voltages applied to respective photodetectors of a BPD \\
  &$i^{(+,-)}$ && Current through respective photodetectors of a BPD \\
  &$i_{max}$ && Maximum current through one of the photodetectors when all input light is directed to that side of the BPD \\
\end{tabularx}
\begin{tabularx}{.8\textwidth}{rrlX}
\multicolumn{4}{c}{\textul{Resolution}} \\
  &$P_{pump}$ && Optical power generated by \textit{a single} pump laser that reaches a corresponding modulator. Does not decribe the wall-plug electrical power used to generate the light \\
  &$P_{las}$ && $=NP_{pump}$. Optical power generated by \textit{all} pump lasers in the system \\
  &{\renewcommand\arraystretch{0.8}$\begin{array}{r} P_{las,thrm-limit}\\P_{las,shot-limit}\\P_{las,rin-limit}\end{array}$} && System-level pump laser power in the thermal, shot, and RIN dominated regimes \\
  &$\eta_{net}$ && Interconnect efficiency due to optical losses following modulation and preceding detection. Has a fixed component and a component proportional to $N$ because the distance traveled through waveguides increases with the physical size of the system. \\
  &$RIN$ && Relative Intensity Noise. The standard deviation of laser output power over average laser power. RIN has a spectrum that varies by laser, but, we approximate it as flat with a typical value of --155~dB/Hz. \\
  &$J^*(B)$ && Proposed variable relating frequency to pump power in the thermal noise dominated regime. It is a function of a fixed photodector impedance \\
  &$E_{thrm}(B)$ && Similar to $J^*(B)$, except it is a function of a fixed photodector capacitance under the assumption that impedance was designed to be optimal for a specific operating frequency. \\
  &$E_{shot}(B)$ && Proposed variable relating frequency to pump power in the shot noise dominated regime. It can be considered a MAC energy limit in some cases. \\
  &$F_{RIN}(B)$ && Proposed variable describing the maximum viable operating frequency for a given resolution requirement. Results from relative intensity noise in a laser. \\
  &$s$ && Normalized cross-correlation of multiple signals after weighting. Describes how power and resolution scale with number of channels. Three special cases of $s$ are shown in Fig.~3, although it can take on continuous values between 0 and 1.
  The $s$ variable plays a role similar to that of the $\rho$ variable introduced in~[45]; however, $\rho$ has an embedded assumption that all signals must be identical (i.e. computation is trivial) in the best-case referred to as ``fixed output precision, only positive inputs/weights''~[45,Table 1]. \\
\end{tabularx}
\begin{tabularx}{.8\textwidth}{rrlX}
\multicolumn{4}{c}{\textul{Cascadability}} \\
  &$g$ && Small-signal input-output gain. It can describe the optical-to-optical gain of a photonic neuron or the voltage-to-voltage gain of an analog photonic link (Note the distinction with $g_{RF}$, which describes the electrical power-to-power gain of an analog photonic link) \\
  &$V_{\pi}$ && Modulator slope efficiency relative to that of a Mach-Zehnder modulator. Used to describe voltage-mode (typically PN junction depletion) modulators, not thermal modulators. The $V_{\pi}$ of a microring modulator is approximately a factor of finesse higher than a Mach-Zehnder based on the same waveguide PN junction. \\
  &$\left.P_{pump}\right|_{g=1}$ && Power required for one neuron reach a unity small-signal gain, i.e. to be cascadable. For a modulator-based O/E/O neuron, it is an optical pump power, but it could refer to electrical power in other types of neurons. \\
  &$E_{aut}$ && Autapse energy. The unity-gain power divided by the frequency of operation. Can be measured using a single neuron with a self connection. \\
  &$P_{las,gain-limited}$ && System-level pump laser power required to meet the gain cascadability condition \\
\end{tabularx}
\begin{tabularx}{.8\textwidth}{rrlX}
\multicolumn{4}{c}{\textul{Appendix: Fundamentals of analog photonic links}} \\
  &$g_{RF}$ && Analog photonic link gain in terms of electrical power ($P=IV$) input to electrical power output \\
  &$SFDR$ && Spurious Free Dynamic Range \\
  &$I_{rec}$ && Average received photocurrent \\
  &$OIP3$ && Output Intercept Point of the third-order harmonic of a non-ideal modulator. Quantifies the effect of modulator nonlinearity on resolution. See~[45] for a full description \\
  &$p_{thrm}$ && Electrical noise power resulting from random motion of electrons in the receiver circuit. Also known as Johnson-Nyquist noise. \\
  &$p_{shot}$ && Electrical noise power resulting from the random process of photoelectric conversion in a photodetector. \\
  &$p_{RIN}$ && Electrical noise power resulting from random fluctuations in pump lasers. \\
  &$p_N (P_N)$ && Total noise power in linear units (decibel units). ``N'' stands for noise, not number of neurons. \\
  &$NF$ && Noise Figure (one variable, not a product of $N$ and $F$) \\
  &$I_d$ && Photodiode dark current \\
  &$K_A$ && Carrier ionization ratio in an avalanche detector \\

\end{tabularx}

\end{document}